\documentclass[preprint,preprintnumbers,nofootinbib,amsmath,amssymb]{revtex4-1}

\usepackage[english]{babel}
\usepackage{color}
\usepackage{amsmath}
\usepackage{physics}
\usepackage{xr-hyper}
\usepackage[hidelinks]{hyperref}
\usepackage{diagbox}
\usepackage{multirow}
\usepackage{amsfonts}
\usepackage{amssymb}
\usepackage{latexsym}
\usepackage{graphicx}
\usepackage{adjustbox}
\usepackage{float}
\usepackage{xr}
\usepackage{caption}
\usepackage{enumitem}
\usepackage[flushleft]{threeparttable}

\usepackage{colortbl}

\usepackage{ytableau}

\usepackage[nodisplayskipstretch]{setspace} 

\usepackage{subfig}

\newcommand{\al}{\alpha}

\newcommand{\pa}{\partial}
\newcommand{\ep}{\epsilon}

\newcommand{\Om}{\Omega}

\newcommand{\de}{\delta}

\newcommand{\De}{\Delta}

\newcommand{\tha}{\theta}

\newcommand{\rar}{\rightarrow}
\newcommand{\lrar}{\leftrightarrow}

\newcommand{\non}{\nonumber}

\begin{document}

\title{Ultra-Compact accurate wave functions for He-like and Li-like iso-electronic sequences and variational calculus. I. Ground state}

\author{A.V.~Turbiner}
\email{turbiner@nucleares.unam.mx, alexander.turbiner@stonybrook.edu}

\author{J.C.~Lopez Vieyra}
\email{vieyra@nucleares.unam.mx}

\author{J.C.~del~Valle}
\email{delvalle@correo.nucleares.unam.mx}

\affiliation{Instituto de Ciencias Nucleares, Universidad Nacional Aut\'onoma de M\'exico,
A. Postal 70-543 C. P. 04510, Ciudad de M\'exico, M\'exico.}

\author{D.J.~Nader}
\email{daniel.nader@correo.nucleares.unam.mx}
\affiliation{ Instituto de Ciencias Nucleares, Universidad Nacional Aut\'onoma de M\'exico, A. Postal 70-543 C. P. 04510, Ciudad de M\'exico, M\'exico }
\affiliation{ Departamento de F\'isica, Universidad Veracruzana, A. Postal 70-543 C. P. 91090, Xalapa, Veracruz, M\'exico. }

\begin{abstract}

Several ultra-compact accurate wave functions in the form of generalized Hylleraas-Kinoshita functions and Guevara-Harris-Turbiner functions, which describe the domain of applicability
of the Quantum Mechanics of Coulomb Charges (QMCC) or, equivalently, the Non-Relativistic QED (NRQED), for the ground state energies (4-5 significant digits (s.d.)) of He-like and Li-like iso-electronic sequences in the static approximation with point-like, infinitely heavy nuclei are constructed. It is shown that for both sequences the obtained parameters can be fitted in $Z$ by simple smooth functions: in general, these parameters differ from the ones emerging in variational calculations. For the He-like two-electron sequence the approximate expression for the ground state function, which provides absolute accuracy for the energy $\sim 10^{-3}$\,a.u. and the same relative accuracies $\sim 10^{-2}-10^{-3}$ for both the cusp parameters and the six expectation values, is found. For the Li-like three-electron sequence the most accurate ultra-compact function taken as the variational trial function provides absolute accuracy for energy $\sim 10^{-3}$\,a.u., 2-3 s.d. for the electron-nuclear cusp parameter for $Z \leq 20$ and 3 s.d. for the two expectation values for $Z=3$.

\end{abstract}

\maketitle
\newpage

\section*{Introduction}

Needless to say that wave functions which are compact, but still quite accurate, are extremely valuable as tools for gaining understanding of quantum systems. It is especially important when these functions describe the domain of applicability of non-relativistic quantum mechanics, the domain which is free of corrections of any type like relativistic etc. It is worth noting that in atomic physics compact accurate functions are of a great importance in study of atomic collisions.
Among compact wave functions one can find those which are the close to the exact wave functions being uniform local approximation of them in the whole coordinate space, thus, implicating the approximate solution of the original quantum problem. Such approximations are quite valuable to find accurately not only the energy but also the expectation values. They can be used as zero approximation in developing convergent perturbation theory in order to increase accuracy \cite{Turbiner:1980-4}. From other side taking linear superpositions of compact functions (with different parameters) as trial function in Rayleigh-Ritz method one can reach high accuracy in energy.

Probably, the idea to look for compact wave functions for atomic systems can be assigned to
E Hylleraas \cite{Hylleraas} who tackled Helium atom in the static approximation, with an infinitely heavy, point-like nucleus (the one-center problem). He had introduced the function, see \cite{Hylleraas}, eq.(13), which was later called the {\it Hylleraas function}, in the form of symmetrized product of three (modified by screening) Coulomb Orbitals,
\begin{equation}
\label{Hylleraas}
  \Psi^{(H)}_0 \ =\ \frac{1}{2}(1 + P_{12})\, e^{- \al\, r_1 - \beta\, r_2 + \gamma\, r_{12}}\ ,
\end{equation}
where $P_{12}$ is the permutation operator of electrons, $r_{1,2,12}$ are electron-nucleus, electron-electron distances and $\al, \beta, \gamma$ are parameters.
Taking (\ref{Hylleraas}) as a trial function for the ground state
with $\al, \beta, \gamma$ playing the role of non-linear variational parameters,
2-5 s.d. - before rounding - in total energies {\it vs} increasing nuclear charge $Z \in [1,20]$ are reproduced, see e.g. \cite{AOP:2019}. In fact, function (\ref{Hylleraas}) with optimal variational parameters $\al_v, \beta_v, \gamma_v$ is one of
the most accurate 3-parametric trial function in domain $Z \in [1,20]$. It describes with high accuracy the domain of applicability of QMCC \footnote{the Quantum Mechanics of Coulomb Charges or, equivalently, the Non-Relativistic QED (NRQED), the theory of Coulomb charges without photons} for He-like systems: He, Li$^+$, Be$^{++}$ etc, and leads to the Majorana-type formula for the ground state energies, see \cite{AOP:2019}.
Taking (\ref{Hylleraas}) as the entry one can construct the effective potential \cite{Turbiner:1980-4},
\begin{equation}
\label{V0}
  V_0\ =\ \frac{\De \Psi^{(H)}_0}{\Psi^{(H)}_0}\ ,
\end{equation}
see for discussion \cite{AOP:2019}, for which $\Psi^{(H)}_0$ is the exact ground state wave
function. It can be easily checked that $V_0$ mimics the original potential $V$, even reproducing the Coulomb singularities. Furthermore for artificially constrained parameters $\al + \beta = Z$ , $\gamma=1$, it leads to the exact values of its residues (cusp parameters). Making a comparison between the original potential $V$ and $V_0$ it can be easily checked that
\[
   \bigg|\frac{V-V_0}{V}\bigg|\ \lesssim \mbox{const}\ ,
\]
is bounded. It guarantees the finite radius of convergence of the perturbation theory in
\begin{equation}
\label{PT}
  V_1 \ =\ V\ -\ V_0\ ,
\end{equation}
- the deviation of the original potential $V$ from $V_0$ effective one
\footnote{In \cite{Turbiner:1980-4} where this perturbation theory was described, it was called the {\it Non-Linearization procedure}. In this procedure the variational energy is given by the sum of the first two terms, $E_v(\al_v, \beta_v, \gamma_v)=E_0+E_1$.} - since $V_1$ is a subdominant perturbation with respect to $V_0$, see e.g. \cite{Berezin-Shubin}.
It is evident that the convergence of the perturbation series occurs for parameters,
\[
   \al - \gamma > 0\ ,\ \beta - \gamma > 0 \ ,
\]
which guarantees the normalizability of (\ref{Hylleraas}) not only for the optimal (variational) parameters $\al_v, \beta_v, \gamma_v$ emerging in variational calculus.

However, after the original work \cite{Hylleraas} further development has mostly been focused on construction of linear combinations of thousands of terms (configurations) as trial wave function \footnote{Thus, realizing the (non)-linear Rayleigh-Ritz method.} in order to reach extreme accuracy \footnote{Sometimes it is well beyond of reach of experiment.} in energy rather than a great compactness and a reasonable accuracy (for discussion see \cite{Harris}). Evidently, in such an approach the difficulty in interpreting the essence of accurate wave functions occurs (as well as in printing thousands of (non)-linear parameters which enter in the above-mentioned linear combinations). Mulliken summarized (and predicted) this situation in the famous phrase \cite{Mulliken} {\it \ldots the more accurate the calculations became, the more the concepts tended to vanish into thin air}. It is worth noting the well-known and evident drawback of this approach: the fast convergence in energy does not guarantee the equally fast convergence in other expectation values.

\medskip

In this paper we return to the original Hylleraas idea of searching compact functions for the He atom in its ground state in a form of single configuration with a few free parameters extending it to He-like two-electronic sequence $(Z;2e)$. In general, the idea to fix free parameters following the minimum of variational energy (the Minimax principle) is abandoned in favor of the idea to find parameters leading to a uniformly-accurate description of all expectation values (including energy) plus getting analytic formulas for parameters as the function of $Z$. However, as the first introductory step, the standard variational calculation should be carried in order to get an understanding on overall quality and adequacy of the considered trial function.

\medskip

It is worth emphasizing that Carter \cite{Carter:1966} and eventually Korobov et al, \cite{Korobov:2000} demonstrated that a linear superposition of the Hylleraas functions
\begin{equation}
\label{Hylleraas-N}
  \Psi^{(N)}\ =\ \sum_{i=1}^N c_i \Psi^{(H)}(r; \al_i, \beta_i, \gamma_i)\ ,
\end{equation}
leads to highly accurate results for energy and the fastest convergence with increase of the length $N$ among different bases commonly used, where both linear parameters $\{c\}$ and non-linear ones $\{\al, \beta, \gamma\}$ are found through a minimization procedure. Furthermore, while minimizing one can see that the parameters $\al_i, \beta_i, \gamma_i$ of the $i$th term $\Psi^{(H)}_i$ demonstrate the dependence on the length $N$ of the linear combination in (\ref{Hylleraas-N}). Hence, the basis is variable: it changes with $N$. The striking fact is that if the first basis function $\Psi^{(H)}_1(r; \al^{(1)}_1, \beta^{(1)}_1, \gamma^{(1)}_1)$ in (\ref{Hylleraas-N}) for $N=1$ is chosen to be optimal, then for larger $N > 1$ the first function $\Psi^{(H)}_1(r; \al^{(N)}_1, \beta^{(N)}_1, \gamma^{(N)}_1)$ as the result of the overall minimization is not anymore optimal: its variational parameters are different from $\al^{(1)}_1, \beta^{(1)}_1, \gamma^{(1)}_1$ but remain close. From the other side, keeping parameters $\al^{(1)}_1, \beta^{(1)}_1, \gamma^{(1)}_1$ in the first function fixed and making minimization with respect to the other parameters deteriorates the variational energy but insignificantly(!). In turn, by analysing the results for energy one can see that the convergence of the perturbation theory with $\Psi^{(H)}_1(r; \al^{(N)}_1, \beta^{(N)}_1, \gamma^{(N)}_1)$ as the zero approximation remains and its rate of convergence depends on $N$ non-essentially
\footnote{Recently, this was checked in detailed study of the cubic, quartic
and sextic radial anharmonic oscillators \cite{delValle}: the first ten figures in energies of several low-lying states are given by the sum of the first three terms $(E_0+E_1+E_2)$ in perturbation theory (in Non-Linearization procedure) {\it independently} on the non-linear parameters in close proximity to optimal (variational) ones of the zero approximation.}.

 \medskip

As for the lithium atom in its ground state an attempt was already made to find a compact function in the form similar to Hylleraas function (\ref{Hylleraas}) taking the (anti)symmetrized product of six (modified by screening) Coulomb Orbitals, see \cite{TGH2009}; recently, it was extended to the Li-like system at $Z > 3$ \cite{AOP:2019}. However, the goal of describing the domain of applicability of QMCC for the Li-like three-electronic sequence was not reached. In the current paper the functions presented in \cite{TGH2009,AOP:2019} are subsequently generalized in order to find one for which the variational energies are able to recover the domain of applicability of QMCC with maximally accurate cusp parameters. Since the Fock expansion for Li-like sequence is not yet constructed, we are unable to construct a uniform approximation of the ground state in the whole configuration space.

The paper consists of Introduction, two Sections and Conclusions, and one Appendix. In Section I the Hylleraas function is revisited as well as the Kinoshita function, they are interpolated and their two subsequent generalizations are proposed with parameters described by simple functions in $Z$. The second generalization is built requiring a uniform description of energy, cusp parameters and available expectation values. In Section II the ultra-compact variational trial function for the Li-like sequence in the spin-doublet state of the lowest energy is found with parameters given by second degree polynomials in $Z$ or in $1/Z^2$. The goal to construct a uniform approximation of the exact function as was made for He-like sequence is not reached for the lithium sequence. In Appendix the results about approximate ground state wavefunction for two-electron system with infinite massive charge $Z$ are collected and formulated in a form of conjecture.

\section{He-like two-electron sequence: ground state}
\vspace{-5mm}
\subsection{Generalities. Hylleraas and Kinoshita functions}
\vspace{-5mm}

It is well known that the ground state for the helium atom is spin singlet with zero total angular momentum (para-Helium), its wavefunction is the product of orbital (coordinate) function on spin function. In general, the orbital function for spin-singlet state with zero total angular momentum depends on relative distances only, see e.g. \cite{twe}, it is symmetric with respect to permutation of electrons,
\[
   \Psi (r_1,r_2,r_{12})\ =\ \Psi (r_2,r_1,r_{12})\ ,
\]
and it obeys to the reduced Schr\"odinger equation
\begin{equation}
\label{SE-radial}
  \hat{H}\,\Psi (r_1,r_2,r_{12})\ =\ E\,\Psi (r_1,r_2,r_{12})\ ,
\end{equation}
where $\hat{H}=\hat{H}(r_1,r_2,r_{12})$,
\begin{gather}
   \hat{H} \ =\ -\frac{1}{2}
\left[
  \frac{\pa^2}{\pa r_1^2}\ +\ \frac{\pa^2}{\pa r_2^2}\ +\ 2\frac{\pa^2}{\pa r_{12}^2}\ +\
         \frac{2}{r_1}\frac{\pa}{\pa r_1}\ +\ \frac{2}{r_2}\frac{\pa}{\pa r_2}\ +\
                     \frac{4}{r_{12}}\frac{\pa}{\pa r_{12}}
\right. \non
\\
\left.
  +\ \left(
  \frac{r_1^2-r_2^2+r_{12}^2}{r_1r_{12}}\right)\frac{\pa^2}{\pa r_1 \pa r_{12}}
  \ +\
  \left(\frac{r_2^2-r_1^2+r_{12}^2}{r_2r_{12}}\right)\frac{\pa^2}{\pa r_2 \pa r_{12}}
    \right]
\non\\
  +\ \left[\frac{-Z}{r_1}\ +\ \frac{-Z}{r_2}\ +\ \frac{1}{r_{12}}\right]\ ,
\label{Hred}
\end{gather}
see \cite{GAM:1987}, Eq.(5), is the so-called {\it radial} Schr\"odinger Hamiltonian \cite{twe}.
As $Z \rar \infty$ for the lowest energy spin-singlet state the exact eigenfunction is known, it is the product of $(1s1s)$ Coulomb orbitals,
\begin{equation}
\label{He-infty-psi}
  \Psi_0\ =\ e^{- \al Z r_1 - \beta Z r_2}\ ,
\end{equation}
at $\al=\beta=1$ with eigenvalue,
\begin{equation}
\label{He-infty-E}
  E_0\ =\ - Z^2  \ .
\end{equation}
In celebrated $1/Z$-expansion \cite{Hylleraas} Eqs.(\ref{He-infty-psi})-(\ref{He-infty-E}) represent zero-order approximation.

The original Hylleraas function (\ref{Hylleraas}) at $Z=2$ can be modified to the case of He-like system $(2e; Z)$ for any integer $Z \geq 1$, indicating explicitly possible $Z$-dependence in parameters,
\begin{equation}
\label{HeZ}
  \Psi_H(r_1,r_2,r_{12})\ =\ \frac{1}{2}(1 + P_{12})\, e^{ - \al Z r_1 - \beta Z r_2 + \gamma r_{12}}\ ,
\end{equation}
with three free parameters $\al, \beta, \gamma$, see \cite{AOP:2019}
\footnote{Present authors are not familiar with previous attempts to use the single Hylleraas function for $Z>2$ (single configuration).}. At the limit $Z \rar \infty$ the parameters $\al,\beta \rar 1$ and $\gamma \rar 0$, the function (\ref{HeZ}) becomes the exact solution (\ref{He-infty-psi}). Long ago it was demonstrated \cite{CL} that all integrals involved to the calculations of the expectation value
of the Hamiltonian are intrinsically 3-dimensional, in particular, in $r_1, r_2, r_{12}$ variables
and they can be evaluated analytically. Eventually, the expectation value of the Hamiltonian
\begin{equation}
\label{H-He}
  {\cal H}\ =\ -\frac{1}{2} \sum_{i=1}^2 \De_i \ -\ \sum_{i=1}^2 \frac{Z}{r_i} \ +\
  \frac{1}{r_{12}}\ ,
\end{equation}
is a certain rational function of parameters $\al, \beta, \gamma$, here $\De_i$ is the Laplacian for the $i$th electron. The procedure of minimization of the expectation value for energy
\begin{equation}
\label{E-exp}
      E_{exp}\ =\ \langle {\Psi_H}|{\cal H}|{\Psi_H} \rangle\ ,
\end{equation}
in order to get the variational energy,
\[
      E_{var}\ =\ {\rm min}_{\al, \beta, \gamma}\ E_{exp}\ ,
\]
is essentially algebraic and can be easily performed. In \cite{AOP:2019}, Fig.{2}, the optimal (variational) parameters $\al_v, \beta_v, \gamma_v$ are presented - they are smooth, slow-changing functions of $Z$. In domain $Z \in [1,20]$ these parameters are fitted with high accuracy by the second degree polynomials in $Z$
\begin{align}
   \al_v\, Z\ &=\ 0.033732\, +\, 1.08564\, Z\, -\, 0.00177\, Z^2 \ ,
\non \\[10pt]
\label{HZ-parms}
  \beta_v\, Z\ &=\ -0.42152\,  +\, 0.926356\, Z\, +\, 0.00131\, Z^2 \ ,
\\[10pt]
   \gamma_v\, Z\ &=\ -0.082824\, +\, 0.247235\, Z\, -\, 0.000125\, Z^2 \ , \non
\end{align}
with very small coefficients in front of $Z^2$ term.
The energy (\ref{E-exp}) obtained with fitted parameters (\ref{HZ-parms}) coincides with both the variational energy and with exact QMCC energies within 4-3 s.d., see Table \ref{H-vs-K}.  The only exception occurs for $Z=1$, where (\ref{E-exp}) with parameters (\ref{HZ-parms}) leads to $-0.523$, the variational energy is $E_{var}=-0.524$, while the exact QMCC energy is $-0.527$ (e.g. \cite{Korobov:2000}). Thus, the trial function (\ref{HeZ}) with parameters (\ref{HZ-parms}) describes the energy in the domain of applicability of QMCC in static approximation, except for $Z=1$. For all $Z \in [1, 20]$ the parameters $\al_v \neq \beta_v$. It indicates to the phenomenon of {\it clusterization}: one electron is closer to the nuclei than another one
\footnote{The importance of the clusterization was observed for the first time when it was found that the Hartree-Fock method does not lead to the bound state of H${}^-$.}.
It measures deviation from the atomic shell model, where both electrons are in $(1s_0)$ state and suppose to be at the same distance to nucleus (in average).

\begin{table}[tb]
\begin{center}
        \caption{The ground state for He-like atoms calculated using the Hylleraas function  $\Psi_H$ (\ref{HeZ}) with parameters (\ref{HZ-parms}) and  the Kinoshita function  $\Psi_K$ (\ref{HeZ-Kinoshita}) with optimal parameters; the nuclear-electron cusp $C_{Z,e}$ and electron-electron  cusp $C_{e,e}$ calculated via expansion (\ref{cusps-2}) (the first lines), and via (\ref{cusp-Ze})-(\ref{cusp-ee}) (the second lines).}
\label{H-vs-K}
               {\setlength{\tabcolsep}{0.45cm} \renewcommand{\arraystretch}{1}
 \resizebox{0.95\textwidth}{!}{%
 \begin{tabular}{c|ccc|ccc}
                        \hline
                        \multirow{2}{*}{$Z$} & \multicolumn{3}{c|}{Hylleraas $\Psi_H$}     &
\multicolumn{3}{c}{Kinoshita $\Psi_K$}    \\
               & $E_{var}$ (a.u.) & $C_{Z,e}$ & $C_{e,e}$ & $E_{var}$ (a.u.) & $C_{Z,e}$ & $C_{e,e}$ \\[4pt]
                        \hline
                        \rule{0pt}{4ex}
                1 & -0.5239   & 0.812    & 0.164    & -0.5259   & 0.776   & 0.313
\\[2pt]
                  &           & 1.030    & 0.164    &           & 0.990   & 0.313
\\[2pt]
                2 & -2.8995   & 1.817    & 0.206    & -2.9014   & 1.822   & 0.293
\\[2pt]
                  &           & 1.954    & 0.206    &           & 1.963   & 0.293
\\[2pt]
               10 & -93.9026  & 9.843    & 0.238    & -93.9032  & 9.838   & 0.254
\\[2pt]
                  &           & 9.962    & 0.238    &           & 9.953   & 0.254
\\[2pt]
               20 & -387.6531 & 19.834   & 0.241    & -387.653  & 19.839  & 0.249
\\[2pt]
                  &           & 19.9453  & 0.241    &           & 19.952  & 0.249
\\[2pt]
\hline
\end{tabular}}}
\end{center}
\end{table}


Employing the Schr\"odinger equation for the Hamiltonian (\ref{H-He}) or (\ref{SE-radial}) one can
construct the Fock expansion of the (exact) ground state function at small distances
\begin{equation}
\label{cusps-2}
  \Psi\ =\ 1\ -\  C_{Z,e}\,(r_1\,+\,r_2)\ +\  C_{e,e}\, r_{12}\ +\ \mbox{\it ``log terms"} \ +\ A\, (r_1^2 + r_2^2)\ +\ B\, r_1 r_2\ +\ldots\ ,
\end{equation}
see e.g. \cite{Myers:1991}, where the coefficients
\[
    C_{Z,e}\ =\ Z\ ,\ C_{e,e}\ =\ \frac{1}{2}\ ,
\]
were found for the first time in \cite{Bartlett:1937}; these coefficients are called the {\it cusp} parameters, they are equal to residues at Coulomb singularities in (\ref{H-He}) at $r_1=0$, $r_2=0$ and $r_{12}=0$, respectively,
{\it ``log terms"} are $o(r^{2-\ep}), \ep >2$. The parameters
\[
     A\ =\ -\frac{E}{6}\ ,\ B \ =\ Z^2 (1 + \frac{1}{3Z})\ ,
\]
are printed in \cite{Myers:1991}\, \footnote{We suspect that the parameter $A$ was printed in \cite{Myers:1991} incorrectly, it should be $A=-\frac{E}{2}$ in agreement with Eq.(\ref{cusps-2-exact}).
}. At the limit $Z \rar \infty$ the exact solution (\ref{He-infty-psi}) generates at small $r_{1,2}$ the double Taylor expansion,
\begin{equation}
\label{cusps-2-exact}
  \Psi\ =\ 1\ -\  Z\,(r_1\,+\,r_2)\ +\ Z^2 (r_1^2 + r_2^2)\ +\ Z^2 r_1 r_2\ +\ \ldots\ ,
\end{equation}
cf. (\ref{cusps-2}),
where the term $r_{12}$ and all "log terms" disappear, $A=-\frac{E}{2}$, see footnote ${}^6$, and $B=Z^2$, here $E=-Z^2$ is the exact eigenvalue.

Taking the Schr\"odinger equation for the Hamiltonian (\ref{H-He}), it can be also derived that
\begin{equation}
\label{cusp-Ze}
 C_{Z,e}\ \equiv\ -\ \frac{\langle \de({\bf r}_i)\frac{\pa}{\pa r_i} \rangle}{\langle
   \de({\bf r}_i) \rangle}\ , \qquad i= 1,2\ ,
\end{equation}
see \cite{CS:1969}, and
\begin{equation}
\label{cusp-ee}
  C_{e,e}\ \equiv\ \frac{\langle \de({\bf r}_{12})\frac{\pa}{\pa r_{12}} \rangle}
  {\langle \de({\bf r}_{12}) \rangle}\ ,
\end{equation}
see also \cite{Harris}, in terms of expectation values without referencing to the Fock expansion at small distances. These formulas should be valid for any multi-electron atom. For a concrete trial function the cusp parameters obtained through its expansion at small distances (\ref{cusps-2}) and those obtained through ratio of expectation values (\ref{cusp-Ze}), (\ref{cusp-ee}) might be {\it different}, see Table I \footnote{It can be shown that accidentally for the Hylleraas and Kinoshita functions the electron-electron cusps obtained via expansion (\ref{cusps-2}) and via the formula (\ref{cusp-ee}) {\it coincide}.}. Needless to say that the cusp parameters obtained through expansion have the meaning of residues of Coulomb singularities in effective potential (\ref{V0}) for which the trial function is exact. Through this text we will rely mainly on those cusp parameters, although the cusp parameters obtained via (\ref{cusp-Ze}), (\ref{cusp-ee}) can give more accurate results like it is in Table I.

Overall quality of the function (\ref{HeZ}) can be ``measured" by how accurately this variational trial function reproduces both the electron-nucleus cusp parameter $C_{Z,e}$ by the expression $\frac{(\al+\beta)Z}{2}$, see Fig.3 in \cite{AOP:2019}, and electron-electron cusp $C_{e,e}$ by $\gamma$ obtained through its expansion at small distances. As for $C_{Z,e}$ at $Z=1$ the difference is of order 20$\%$, then it reduces to 1$\%$ at $Z=12$ and tends to zero at larger $Z$. In general, $C_{Z,e}=\frac{(\al_v+\beta_v)Z}{2}$ grows with increase $Z$ for the function (\ref{HeZ}).
As for the $C_{e,e}=\gamma$ found by using the function (\ref{HeZ}), it grows with increase of $Z$ but not well-reproduced by the parameter $\gamma_v$. For large $Z$ it differs in $\gtrsim 50\%$ systematically while for small $Z$ the difference can reach several hundred percents
\footnote{It is well known fact about difficulty to reproduce the electron-electron cusp parameter in variational calculation, see e.g. \cite{Harris}.}
. It means that the Hylleraas function $\Psi_H(r; \al_v, \beta_v, \gamma_v)$ does not behave correctly in vicinity of $r_{12}=0$; at the same time, it does not influence the quality of variational energy, hence, the contribution to variational integrals coming from this domain is non-essential. However, if the parameter $\gamma$ in (\ref{HeZ}) is fixed manually to be equal 1/2 in order to reproduce $C_{e,e}$ exactly, it leads to a significant deterioration of the variational energy, specially at small $Z$:  for $Z=1$ it goes from $\simeq -0.477$ (thus, no binding occurs for H${}^-$(!)) and $C_{Z,e}\sim 1.074$, hence, being $\sim 10\%$ of relative difference in energy from the exact value, then at $Z=2$ to $\sim -2.858$ and $C_{Z,e} \sim 2.037$ ($\sim 2\%$ of relative difference in the energy from the exact value) and then goes down to $\sim -93.87$ and $C_{Z,e} \sim 10.0$ for $Z=10$ (with a relative difference of $\sim 0.04\%$ from the exact energy) and eventually to $E_{exp} \sim -387.62$ and $C_{Z,e} \sim 20.0$ at $Z=20$ (with a relative difference in energy $\sim 0.01\%$ from the exact value). All that can be considered as a hint for searching a modification of the function (\ref{HeZ}) which would lead to accurate expectation values and cusp parameters.

\enlargethispage{\baselineskip}
In expansion of the Hylleraas function at small distances the "log terms", see (\ref{cusps-2}), are not generated but quadratic ones in $r_{1,2}$ do,
\[
    \Psi_H\ =\ \ldots \ +\ \frac{Z^2}{4}\,(\al^2+\beta^2) (r_1^2+r_2^2)\ +\ \ldots
\]
It is interesting to compare them with ones which appear in Fock expansion (\ref{cusps-2}),
\[
  A_H\ =\ \frac{Z^2}{4}\,(\al_v^2+\beta_v^2)\ ,
\]
see Table \ref{Table-2}, first rows.
\begin{table}[tb]
\begin{center}
\caption{Normalized electron-nuclear cusp ${\cal C}_{Z,e}= C_{Z,e}/Z$,
         electron-electron cusp $C_{e,e}$ and parameter $A$, see (\ref{cusps-2}).
         First rows: Hylleraas function $\Psi_H$,
         Second rows: Hylleraas-Kinoshita function $\Psi_{HK}$,
         Third rows: the function $\Psi_F$,
         Fourth row: $A_{exact} = -\frac{E}{2}$. }
         \label{Table-2}
 {\setlength{\tabcolsep}{0.75cm} \renewcommand{\arraystretch}{1}
 \resizebox{0.65\textwidth}{!}{%
\begin{tabular}{|c|c|c|c|}
\hline
$Z$ & ${\cal C}_{Z,e}$\ &\ $C_{e,e}$ &  $A$     \\
\hline\hline
1   & \ 0.779 \   & \ 0.147 \ & 0.376    \\
    &  0.776      &  0.313    & 0.414    \\
    &  0.906      &  0.453    & 0.477    \\
    &      1      &  1/2      & 0.264    \\[4pt]
\hline
2   &  0.912      & 0.207     & 1.724   \\
    &  0.911      & 0.293     & 1.801   \\
    &  0.957      & 0.478     & 1.930   \\
    &             &           & 1.452   \\[4pt]
\hline
 10 &  0.984      & 0.237     & 48.822  \\
    &  0.984      & 0.254     & 49.368  \\
    &  0.990      & 0.495     & 49.487  \\
    &             &           & 46.953  \\[4pt]
\hline
20  &  0.992      & 0.241     & 197.421  \\
    &  0.992      & 0.249     & 199.187  \\
    &  0.995      & 0.498     & 198.859  \\
    &             &           & 193.829  \\[4pt]
\hline\hline
\end{tabular}}}
\end{center}
\end{table}
As an alternative to (\ref{HeZ}), Kinoshita \cite{Kinoshita} proposed the function
\begin{equation}
\label{HeZ-Kinoshita}
  \Psi_K(r_1,r_2,r_{12}) \ =\ \frac{1}{2}(1 + P_{12})\,\bigg[\,(1 + b r_{12})\, e^{ - \al Z r_1 - \beta Z r_2}\,\bigg]\ ,
\end{equation}
where $b,\al,\beta$ are parameters.
It turned out that being taken as the trial function the Kinoshita function (\ref{HeZ-Kinoshita}) is the most accurate 3-parametric trial function in domain $Z \in [1,20]$. It gives the better results for variational energy $E_{var}$ and electron-electron cusp $C_{e,e}$ than those based on the Hylleraas function (\ref{HeZ}), being slightly worse
as for electron-nuclear cusp $C_{Z,e}$, see Table \ref{H-vs-K}. Straightforward attempt to impose constraints $b=1/2$ and $(\al+\beta)=1$ to get exact values for cusps leads to a dramatic deterioration of variational energies similar to what happened for the Hylleraas function $\Psi_H$ (\ref{HeZ}). Serious drawback of the Kinoshita function is of qualitative nature: the electron-electron cusp $C_{e,e}$ found with (\ref{HeZ-Kinoshita}) decreases with increase of $Z$. One can draw a conclusion that the Kinoshita function should be modified in order to change this behavior.

\enlargethispage{\baselineskip}

\subsection{Interpolating Hylleraas and Kinoshita functions}
Natural interpolation of the Hylleraas (\ref{HeZ}) and Kinoshita (\ref{HeZ-Kinoshita}) functions is of the form
\begin{equation}
\label{HeZ-mod}
  \Psi_{HK}\ =\ \frac{1}{2}(1 + P_{12})\,[(1 - a r_1 + b r_{12})\,
  e^{ - \al Z r_1 - \beta Z r_2 + \gamma r_{12}}]\ ,
\end{equation}
where $a, b, \al, \beta, \gamma$ are parameters. It coincides with the Hylleraas function at $a=b=0$, while at $a=\gamma=0$ it is reduced to the Kinoshita function. Furthermore, at $Z \rar \infty$ the parameters $a,b, \gamma \rar 0$ and $\al, \beta \rar 1$, the function $\Psi_{HK}$ becomes the exact ground state function of two non-interacting Hydrogen atoms.

In order to find parameters $a, b, \al, \beta, \gamma$ in (\ref{HeZ-mod}) we impose the following three conditions:

\noindent
(i) the energy $E_{exp}=\langle{\cal H}\rangle|_{\Psi_{HK}}$ should reproduce not less than 4 s.d. for $Z \in [2, 10]$ and not less than 5 s.d. for $Z \in [11, 20]$ before rounding and,

\noindent
(ii) it should lead to maximally accurate value of the electron-electron cusp $C_{e,e}$
given by $(\gamma + b)$, being as close as possible to the exact value 1/2 (\ref{cusps-2}).

\noindent
(iii) in domain $Z\in[1,20]$ the parameters $a, b, \al, \beta, \gamma$ should be fitted by as simple as possible functions in $Z$.

\noindent

It is evident that the expectation value $E_{exp}$ (\ref{E-exp}) over the function $\Psi_{HK}$ (\ref{HeZ-mod}) is rational function of the parameters $a, b, \al, \beta, \gamma$, see \cite{CL}, while $C_{Z,e}$ and $C_{e,e}$ are linear functions, it simplifies analysis.
All three conditions can be accomplished and the resulting parameters are the second degree polynomials in $Z$,
\begin{equation}
\label{parametersHe}
\begin{array}{r@{}l}
 a_{HK}          &{}=\quad  0.16099 + 0.05000\,Z - 0.00065\,Z^2\ , \\[5pt]
 b_{HK}\,Z       &{}=      -0.23516 + 0.59820\,Z + 0.06341\,Z^2\ , \\[5pt]
 \al_{HK}\,Z     &{}=      -0.04841 + 1.05701\,Z - 0.00187\,Z^2\ , \\[5pt]
 \beta_{HK}\,Z   &{}=      -0.44227 + 0.90626\,Z + 0.00231\,Z^2\ , \\[5pt]
 \gamma_{HK}\,Z  &{}=\quad  0.25820 - 0.20821\,Z - 0.05817\,Z^2\ ,
\end{array}
\end{equation}
cf.(\ref{HZ-parms}).
%
Making comparison of the  parameters (\ref{parametersHe}) of the modified function (\ref{HeZ-mod}) with the variational parameters found for the Hylleraas function, see (\ref{HZ-parms}). One can find out that the parameters $\al_{HK}\, Z$ and $\beta_{HK}\, Z$ in (\ref{HeZ-mod}) behave in similar way as the variational parameters $\al_v Z$ and $\beta_v Z$ in Hylleraas function. In general, they have to approximate almost linear behavior as a function of $Z$: the parameters $\al_{HK} Z$ and $\al_v Z$ deviate slightly for large $Z$, while the parameters $\beta_{HK} Z$ and $\beta_v Z$ almost coincide at $Z \in [2,20]$. In turn, the parameters $\gamma_{HK} Z$ and $\gamma_v Z$ behave  differently: while the variational parameter $\gamma_v Z$ exhibits the linear growth with $Z$, the parameter $\gamma_{HK} Z$ in (\ref{HeZ-mod}) has a pronounced quadratic in $Z$ behavior.

\bigskip

These parameters inserted into the function (\ref{HeZ-mod}) allow to obtain the electron-nuclear cusp in the form
\begin{equation}
\label{cuspkato:Ze}
  C_{Z,e}^{(HK)}\ =\  \frac{(\al_{HK}\ +\ \beta_{HK})\,Z\,+\,a_{HK}\,}{2}\ =\
  -\ 0.164843\ +\ 1.00664\,Z\ -\ 0.00011\,Z^2 \quad ,
\end{equation}
and also the electron-electron cusp,
\begin{equation}
\label{cuspkato:ee}
\quad
C_{e,e}^{(HK)}\ =\ \gamma_{HK}\ +\ b_{HK} \ =\ 0.02306\,Z^{-1}\ +\ 0.38998\ +\ 0.00524\,Z \quad ,
\end{equation}
which ranges from $C_{e,e} \sim 0.41$ for $Z = 2$ to $\sim 0.50$ for $Z = 20$. Simultaneously,
the energy follows the Majorana formula \cite{AOP:2019},
\begin{equation}
\label{majorana-2}
  E^{(2)}_{HK}(Z)\ =\ -\,0.153282\ +\ 0.624583\, Z\ -\,Z^2\ ,
\end{equation}
see Table \ref{TABLE-3}. It also leads to the parameter $A$ in (\ref{cusps-2}) in the form
\[
    A_{HK}\ =\ \frac{Z^2}{4}\,(\al_{HK}^2+\beta_{HK}^2)\ +\ \frac{Z}{2}\, a_{HK}\,\al_{HK}\ ,
\]
see Table \ref{Table-2}, second row.

\bigskip

Using the function $\Psi_{HK}$ (\ref{HeZ-mod}) with parameters (\ref{parametersHe}) we also calculated some expectation values (which are known to be the most difficult to find accurately), see Table \ref{expectation1}. For $Z \in [1, 10]$ the agreement with previous calculations \cite{Harris,Drake,Frolov} \footnote{Note the results from \cite{Harris,Drake,Frolov} coincide in 3-4 s.d.} is 1-2 s.d. except for $\langle\bf{r}_1\cdot\bf{r}_2\rangle$ at $Z=1$ where the difference reaches $\sim 40\%$ \footnote{
Note taking instead of the parameters (\ref{parametersHe}) the variational (optimal) parameters in $\Psi_{HK}$ for $Z=1$, this expectation value is improved dramatically: agreement becomes to $\sim 1\%$, while deteriorating the agreement in other expectation values}.

\bigskip

\noindent
{\it Perturbation Theory in Non-Linearization procedure \cite{Turbiner:1980-4}}\ .
Knowledge of the parameters (\ref{parametersHe}) in $\Psi_{HK}$ (\ref{HeZ-mod}) allows us to calculated the effective potential (\ref{V0}),
\begin{equation}
\label{VM}
  V_{\rm eff}\ =\ \frac{\De \Psi_{HK}}{\Psi_{HK}}\ ,
\end{equation}
for which $\Psi_{HK}$ is the exact ground state wave function and then develop perturbation theory in difference
\[
   V_{PT}\ =\ (V - V_{\rm eff})\ .
\]

It can be easily checked that $V_{PT}$ is subordinate being bounded perturbation
with respect to the original potential $V$ and the resulting perturbation theory
should be convergent. Knowing the expectation value
$$E_{exp}\ =\ \langle {\Psi_{HK}}|{\cal H}|{\Psi_{HK}} \rangle\ ,$$ cf.(\ref{E-exp}),
one can calculate the first correction $\de \phi_{HK}$ to the exponential phase $\phi_{HK}$:
$(\phi_{HK}+\de \phi_{HK})$ of the wavefunction $\Psi_{HK}$:
\[
      \Psi_{HK}\ =\ e^{{- \phi_{HK} }}\ .
\]
It obeys the equation
\begin{equation}
\label{de-phi}
    (\De_{\bf r_1} +  \De_{\bf r_2}) \de \phi_{HK} -
    2 (\nabla_{\bf r_1}\phi_{HK} \cdot \nabla_{\bf r_1} \de \phi_{HK}) -
    2 (\nabla_{\bf r_2}\phi_{HK} \cdot \nabla_{\bf r_2} \de \phi_{HK})\ =\ E_{exp}\,-\,V_{PT}\ ,
\end{equation}
see \cite{Turbiner:1980-4}. It allows us to construct the corrected wavefunction,
\begin{equation}
\label{PsiM-c}
     \Psi_{HK}^{(c)}\ =\ \Psi_{HK}\, (1 - \de \phi_{HK})\ .
\end{equation}
With this function one can calculate the first correction to expectation values,
\begin{equation}
\label{O-exp}
      O_{exp}\ =\ \langle {\Psi_{HK}}|{\cal O}|{\Psi_{HK}} \rangle \ -
      \ 2 \langle {\Psi_{HK}}|{\cal O}|{\Psi_{HK} \cdot \de \phi_{HK}}\rangle \ ,
\end{equation}
neglecting the contribution $\sim (\de \phi_{HK})^2$. In particular, using this formula one can make the estimate of order of the correction to the energy. It is
of the order of 4th s.d. for \hbox{$Z \leq 10$} and of the order 5th s.d. for~\hbox{($Z \geq 10$)}.
It seems interesting to make analysis of permutationally anti-symmetric function
\begin{equation}
\label{HeZ-mod-A}
  \Psi^{(-)}_{HK}\ =\ \frac{1}{2}(1 - P_{12})\,\big[(1 - a r_1 + b r_{12})\,
  e^{ - \al Z r_1 - \beta Z r_2 + \gamma r_{12}}\big]\ ,
\end{equation}
cf.(\ref{HeZ-mod}), to study the lowest energy spin-triplet state (ortho-Helium).
It will be carried out elsewhere.

\vskip -1.5cm

\subsection{Generalized Hylleraas-Kinoshita function}

\vskip -0.5cm

Even though trial functions (\ref{HeZ}) and  (\ref{HeZ-mod}) describe with high accuracy the energy of the ground state for He-like two-electron systems in the domain of applicability of QMCC (and beyond) these functions do not describe accurately the region of small inter-electronic distances. It is well seen in a description of electron-electron cusp (see Table \ref{H-vs-K} and \ref{TABLE-3}, first rows).  Even further improvement of the trial function (\ref{HeZ-mod}) in ``$r_{12}$ direction" is needed to describe the linear behavior of the phase at large and small inter-electronic distances but with different slopes~\footnote{From physics point of view it means no screening at small distances (pure Coulomb repulsion of electrons) occurs while a screening at large distances due to presence of nuclei ``in between" the electrons does occur.}. Simple interpolation consists in making the screening parameter $\gamma$ dependable on $r_{12}$,
\begin{equation}
\label{He-rational}
    \gamma \to \gamma \frac{(1 + c\, r_{12})}{(1 + d\,r_{12})}\ ,
\end{equation}
where $c,d$ are parameters, see e.g. \cite{BMBM:2001}. This replacement effectively interpolates the value of the screening parameter $\gamma$ at small $r_{12}$ to $\gamma c/d$ at large $r_{12}$. This replacement leads to 7-parametric function of the form
\begin{equation}
\label{psinew}
\Psi_F\ =\ \frac{1}{2}\,(1 + P_{12})\,\bigg[
(1 - a r_1 + b r_{12})\,e^{- \al Z r_1\, -\, \beta Z r_2\, +\, \gamma r_{12}\, \frac{(1 + c r_{12})}{(1 + d r_{12})}}\bigg]\ ,
\end{equation}
where $a, b, c, d, \al, \beta, \gamma$ are parameters, $P_{12}$ is permutation operator. Evidently, for $c=d=0$  the function $\Psi_F$ degenerates to $\Psi_{HK}$. Formulas for cusps in terms of parameters of the function (\ref{psinew}) remain the same as for $\Psi_{HK}$ (\ref{HeZ-mod}), see (\ref{cuspkato:Ze}), (\ref{cuspkato:ee}).

\vfil

\begin{table}[H]
                \caption{\footnotesize
        Ground state energy $E$ for the He-like sequence
        calculated with function $\Psi_{HK}$ (\ref{HeZ-mod}) and parameters (\ref{parametersHe}) (first row), and with function $\Psi_F$ (\ref{psinew}) constrained to (\ref{HK:1 to 2}) (second row) and with parameters (\ref{parametersnew}) constrained to (\ref{1 to 2}) (third row), see text; $E_{exact}$ (rounded to 4 d.d.)
        from \cite{AOP:2019} (and Refs therein).
        Absolute deviation $\De E=E-E_{exact}$, correlation energy $E_{corr}$ and relative deviation $\De E/E_{corr}$ shown and also normalized electron-nuclear cusp
        ${\cal C}_{Z,e}=C_{Z,e}/Z$ and electron-electron cusp $C_{e,e}$.}
\label{TABLE-3}
\begin{center}
{\setlength{\tabcolsep}{0.5cm} \renewcommand{\arraystretch}{1}
 \resizebox{0.8\textwidth}{!}{%
\begin{tabular}{cccccccc}
\hline
$Z$&$E_{exact} (a.u.)$ &\ $E (a.u.)$ &\ $\De E (a.u.)$ &\ $E_{corr} (a.u.)$ &\ $\De E/E_{corr}$ &\
  ${\cal C}_{Z,e}\, (=1)$ &\ $ \,C_{e,e}\, (=1/2)$  \\
\hline
\rule{0pt}{4ex}
1  & -0.5278   & -0.524      &            &           &          &  0.84168  &
         0.418   \\
   &           & -0.52526    &            &           &          &  0.84168  &
         0.42084 \\ 
   &           & -0.52685    &  0.00096   & 0.47316   & 0.00202  & 0.906     & 0.453
    \\[3pt]
2  & -2.9037   & -2.903      &            &           &          &  0.924    &
         0.412   \\
   &           & -2.9027     &            &           &          &  0.924    &
         0.462   \\ 
   &           & -2.9026     &  0.00108   & 1.09738   & 0.00099  & 0.95675   &
         0.478375  \\[3pt]
3  & -7.2799   & -7.279      &            &           &          &  0.95137  &
         0.413    \\
   &           & -7.2787     &            &           &          &  0.95137  &
         0.475685 \\ 
   &           & -7.2788     &  0.00113   & 1.72123   & 0.00066  & 0.97044   &
         0.48522  \\[3pt]
4  & -13.6555  & -13.65      &            &           &          &  0.965    &
         0.417    \\
   &           & -13.6543    &            &           &          &  0.965    &
         0.4825   \\ 
   &           & -13.6544    &  0.00113   & 2.34563   & 0.00048  &  0.97734  &
         0.48867  \\[3pt]
5  & -22.0309  & -22.03      &            &           &          &  0.97314  &
         0.420    \\
   &           & -22.0297    &            &           &          &  0.97314  &
         0.48657  \\ 
   &           & -22.0297    & 0.00116    & 2.97026   & 0.00039  &  0.98158  &
         0.49079  \\[3pt]
6  & -32.4062  & -32.40      &            &           &          &  0.97853  &
         0.425    \\
   &           & -32.4050    &            &           &          &  0.97853  &
         0.489265 \\ 
   &           & -32.4050    & 0.00120    & 3.59500   & 0.00033  &  0.98447  &
          0.492235 \\[3pt]
7  & -44.7814  & -44.78      &            &           &          &  0.98235  &
          0.430    \\
   &           & -44.7803    &            &           &          &  0.98235  &
          0.491175 \\ 
   &           & -44.7802    & 0.00119    & 4.21979   & 0.00028  &  0.98657  &
          0.493285 \\[3pt]
8  & -59.1566  & -59.15      &            &           &          &  0.98519  &
          0.435    \\
   &           & -59.1554    &            &           &          &  0.98519  &
          0.492595 \\ 
   &           & -59.1554    & 0.00123    & 4.84463   & 0.00025  &  0.98817  &
          0.494085  \\[3pt]
9  & -75.5317  & -75.53      &            &           &          &  0.98737  &
          0.440    \\
   &           & -75.5305    &            &           &          &  0.98737  &
          0.493685 \\ 
   &           & -75.5305    & 0.00123    & 5.46953   & 0.00023  &  0.98942  &
          0.49471  \\[3pt]
10 & -93.9068  & -93.90      &            &           &          &  0.9891   &
          0.445    \\
   &           & -93.9055    &            &           &          &  0.9891   &
          0.49455 \\ 
   &           & -93.9055    & 0.00129    & 6.09449   & 0.00021  & 0.99044   &
          0.49522  \\[3pt]
11 & -114.2819 & -114.28     &            &           &          &  0.9905   &
          0.450      \\
   &           & -114.2805   &            &           &          &  0.9905   &
          0.49525   \\ 
   &           & -114.2805   & 0.00141    & 6.71951   & 0.00021  &  0.99127  &
          0.495635  \\[3pt]
12 & -136.6569 & -136.65     &            &           &          &  0.99164  &
          0.455    \\
   &           & -136.6555   &            &           &          &  0.99164  &
          0.49582  \\ 
   &           & -136.6554   & 0.00147    & 7.34457   & 0.00020  &  0.99197  &
          0.495985  \\[3pt]
13 & -161.0320 & -161.03     &            &           &          &  0.99259  &
          0.460    \\
   &           & -161.0305   &            &           &          &  0.99259  &
          0.496295 \\ 
   &           & -161.0303   & 0.00166    & 7.96966   & 0.00021  &  0.99257  &
          0.496285  \\[3pt]
14 & -187.4070 & -187.40     &            &           &          &  0.99339  &
          0.465    \\
   &           & -187.4055   &            &           &          &  0.99339  &
          0.496695 \\ 
   &           & -187.4052   & 0.00176    & 8.59476   & 0.00021  &  0.99309  &
          0.496545  \\[3pt]
15 & -215.7821 & -215.78     &            &           &          &  0.99407  &
          0.470    \\
   &           & -215.7804   &            &           &          &  0.99407  &
          0.497035 \\ 
   &           & -215.7801   & 0.00196    & 9.21986   & 0.00021  &  0.99353  &
          0.496765  \\[3pt]
16 & -246.1571 & -246.15     &            &           &          &  0.99465  &
          0.475    \\
   &           & -246.1554   &            &           &          &  0.99465  &
          0.497325 \\ 
   &           & -246.1551   & 0.00204    & 9.84494   & 0.00021  &  0.99393  &
          0.496965  \\[3pt]
17 & -278.5321 & -278.53     &            &           &          &  0.99515  &
          0.480    \\
   &           & -278.5303   &            &           &          &  0.99515  &
          0.497575 \\ 
   &           & -278.5300   & 0.00210    & 10.47000  & 0.00020  &  0.99428  &
          0.49714  \\[3pt]
18 & -312.9071 & -312.90     &            &           &          &  0.99559  &
          0.486    \\
   &           & -312.9051   &            &           &          &  0.99559  &
          0.497795 \\ 
   &           & -312.9050   & 0.00212    & 11.09502  & 0.00019  &  0.99459  &
          0.497295  \\[3pt]
19 & -349.2822 & -349.28     &            &           &          &  0.99596  &
          0.491    \\
   &           & -349.2799   &            &           &          &  0.99596  &
          0.49798  \\ 
   &           & -349.2800   & 0.00220    & 11.72000  & 0.00019  &  0.99486  &
          0.49743  \\[3pt]
20 & -387.6572 & -387.65     &            &           &          &  0.99629  &
          0.496    \\
   &           & -387.6546   &            &           &          &  0.99629  &
          0.498145 \\ 
   &           & -387.6551   & 0.00214    & 12.34494  & 0.00017  & 0.99511   &
          0.497555  \\[3pt]
 \arrayrulecolor{black}\hline
\end{tabular}}}
\end{center}
\end{table}

As the first step we consider (\ref{psinew}) as trial function, where all 7 free parameters $a, b, c, d, \al, \beta, \gamma$ are variational parameters. It leads for helium atom ($Z=2$) to a certain improvement in the variational energy $E=-2.9032$\,a.u. and the electron-nuclear cusp $C_{Z,e}= 1.8602$, but at the same time, the electron-electron cusp significantly deteriorates, $C_{e,e} = 0.3643$ with respect to $\Psi_{HK}$, cf. Table \ref{TABLE-3}.
This result indicates that the behavior of the exact function at small inter-electron distances
is still not well reproduced within variational calculus: the energy is not that sensitive to the behavior of the wave function in the domain near Coulomb singularity at $r_{12}=0$.

Alternatively, one can require that the function (\ref{psinew}) reproduces the exact values of both cusps taking the relations (\ref{cuspkato:Ze}) and (\ref{cuspkato:ee}) as constraints and treating the remaining 5 parameters as {\it true} variational. As for helium atom, $Z=2$, such an approach turns to be excessively restrictive: it leads to unsatisfactory variational energy $E \sim -2.87$ a.u., cf. Table \ref{TABLE-3}.

Let us assume that relative local accuracy of the approximate wave function with respect to the exact function is reflected in relative accuracies of both cusp parameters.
The function $\Psi_{HK}$ leads already to sufficiently high relative accuracy in electron-nuclear cusp $C^{(HK)}_{Z,e}$ for all studied values of $Z \in [1,20]$. A natural question to ask: {\it Can the same relative accuracy be reached for electron-electron cusp ${\tilde C}^{(F)}_{e,e}$, hence,
\begin{equation}
\label{HK:1 to 2}
  {\tilde C}^{(F)}_{e,e}\ =\ \frac{1}{2Z}\,C^{({HK})}_{Z,e} \ ,
\end{equation}
keeping the same (or higher) relative accuracy in energy $E$, which correspond to the domain of applicability of QMCC?} Keeping $\al=\al_{{HK}}, \beta=\beta_{{HK}}, a=a_{{HK}}$, which guarantee that $C^{({HK})}_{Z,e}$ is unchanged, see (\ref{cuspkato:Ze}), and making constrained variational calculations with
(\ref{HK:1 to 2}) demonstrate easily that answer is affirmative for all $Z \in [1,20]$, see in Table \ref{TABLE-3}, the second rows for each $Z$.
This result leads naturally to the next question: {\it What is maximal relative accuracy which can be reached with function (\ref{psinew}) while both cusps constrained to
\begin{equation}
\label{1 to 2}
  {C}^{(F)}_{e,e}\ =\ \frac{1}{2Z}\,C^{({F})}_{Z,e} \ ,
\end{equation}
with the same (or higher) relative accuracy in electronic correlation energy \footnote{{\it de facto} we require the absolute deviation in energy should be $\sim (1-2) \times 10^{-3}$ for all $Z\in[1,20]$}, which is the difference between total energy and sum of the ground state energies of two $Z$-Hydrogen atoms:}
\begin{equation}
\label{Ecorr}
  E_{corr}\ =\ E\ +\ Z^2\ (\approx\ -\,0.153282\ +\ 0.624583\, Z)\ ,
\end{equation}

%
%
\begin{table}[H]
\caption{Expectation values (in a.u.) for the He-like sequence
  calculated with function $\Psi_{HK}$ (\ref{HeZ-mod}) and parameters (\ref{parametersHe}) (first rows), and with function $\Psi_F$ (\ref{psinew})
  and parameters (\ref{parametersnew}) (second rows). Results labeled by $^{a}$ \cite{Harris},  $^{b}$ \cite{Drake},  $^{c}$ \cite{Frolov},
  $^{d}$ \cite{ThakkarSmith:77} ,  $^{e}$  \cite{Pekeris:58},  $^{f}$ \cite{Nakashima:2008}.}
\label{expectation1}
 \begin{minipage}{\textwidth}
 \begin{threeparttable}
\begin{center}
{\setlength{\tabcolsep}{0.65cm} \renewcommand{\arraystretch}{1.1}
 \resizebox{0.95\textwidth}{!}{%
\begin{tabular}{clllllll}
\hline
\rule{-4pt}{4ex}
$Z$ & $\langle\bf{r}_1\cdot\bf{r}_2\rangle$ & $\langle {r}_{12}\rangle$ & $\langle\delta(\bf{r}_1)\rangle$ & $\langle\de(\bf{r}_{12})\rangle$ & $\langle\delta(\bf{r}_1)\de(\bf{r}_2)\rangle$
& $\langle\de(\bf{r}_1)\frac{\pa}{\pa r_1}\rangle$ &\ $\langle\de(\bf{r}_{12})\frac{\pa}{\pa r_{12}}\rangle$ \\[3pt]
\hline
\rule{-4pt}{5ex}
1   & -1.064                &                      &  0.162                                  &  0.002                               &  0.003             &                           &
\\
    & -0.777\,1             & 4.261\,2             &  0.165\,4                               &  0.002\,28                           &  0.005\,1          & -0.164\,3                 &\  0.002\,0
\\
    & -0.687\,3\,$^{b,c}$   & 4.412\,7\,$^{b,d,f}$ &  0.164\,6\,$^{b,c,}$\ \tnote{$\dagger$} & 0.002\,74\,$^{b,c,d,e}$              &  0.005\,1\,$^{c}$  & -0.164\,6\,$^c$           &   0.001\,4$^c$
\\[2pt]
\rule{-4pt}{4ex}
2   & -0.061                &                      &  1.77                                   & 0.102
    &  1.7                  &                      &
\\
    & -0.071\,1             & 1.426\,8             &  1.782\,5                               & 0.097\,0                             &  1.806             &  -3.509\,3                &\ 0.070\, 6   \\
    & -0.064\,7\,$^{a,b,c}$ & 1.422\,1\,$^{b,d,f}$ & 1.810\,4\,$^{b,c}$                      & 0.106\,3\,$^{b,c}$                   &  1.869\,$^{c}$     &  -3.620\,9\,$^{c}$        & 0.053\,2$^{c}$
\\[2pt]
\rule{-4pt}{4ex}
3   & -0.016                &                      & 6.75                                    & 0.52                                 &  30.2              &                           &                   \\
    & -0.018\,3             & 0.861\,3             & 6.783                                   & 0.511\,2                             &  32.762            & -20.074                   &\ 0.346\, 2   \\
    & -0.017\,3\,$^b$       & 0.862\,3\,$^{b,d,f}$ & 6.852\,$^{a,b,c,d,e}$                   & 0.533\,7\,$^{b,}$\ \tnote{$\dagger$} &                    &   \phantom{-20.557$^{d}$} &  \phantom{0.265\,8$^{d}$}
\\[2pt]
4   & -0.006\,9             &                      &  17.03                                  & 1.49                                 &  220               &                           & \\
    & -0.007\,13            &   0.617\,2           &  17.091                                 & 1.485                                &  228.6             &-67.626                    &\ 0.970\,6    \\
    & -0.006\,91 $^a$       &   0.618\,8\,$^{f}$   &  17.20 $^a$                             & 1.52 $^a$                            & 231 $^a$           &                           & \\[4pt]
5   & -0.003\,55            &                      &  34.53                                  & 3.25                                 & 958                &                           & \\
    & -0.003\,48            &  0.481\,2            &  34.616                                 & 3.254                                & 987.5              & -171.582                  &\ 2.081\,6    \\
    &                       &  0.482\,4\,$^{f}$    &  34.76 $^a$                             & 3.31 $^a$                            & 996 $^a$           &                           & \\[4pt]
6   & -0.002\,08            &                      &  61.19                                  & 6.03                                 & 3\,121             &                           & \\
    & -0.001\,969           &  0.394\,5            &  61.257                                 & 6.052                                & 3\,195.2           & -364.959                  &\ 3.814\,4    \\
    &                       &  0.395\,3\,$^{f}$    &  61.44 $^a$                             & 6.14 $^a$                            & 3\,221 $^a$        &                           & \\[4pt]
7   & -0.001\,32            &                      &  98.94                                  & 10.1                                 & 8\,369             &                           & \\
    & -0.001\,232           &  0.334\,4            &  98.910                                 & 10.11                                & 8\,521.7           & -688.361                  &\ 6.302\,1   \\
    &                       &  0.334\,8\,$^{f}$    &  99.16 $^a$                             & 10.3 $^a$                            & 8\,596 $^a$        &                           & \\[4pt]
8   & -0.000\,89            &                      &  149.69                                 & 15.6                                 & 19\,527            &                           & \\
    & -0.000\,830           &  0.290\,2            &  149.470                                & 15.67                                & 19789.4            & -1189.98                  &\ 9.676\,0   \\
    &                       &  0.290\,4\,$^{f}$    &  149.83 $^a$                            & 15.9 $^a$                            & 19\,975 $^a$       &                           & \\[4pt]
9   & -0.000\,63            &                      &  215.37                                 & 22.9                                 & 40\,843            &                           & \\
    & -0.000\,591           &  0.256\,4            &  214.836                                & 22.96                                & 41\,414.7          & -1925.61                  &\ 14.066\, 1 \\
    &                       &  0.256\,4\,$^{f}$    &  215.34 $^a$                            & 23.3 $^a$                            & 41\,750 $^a$       &                           & \\[4pt]
\rule{-4pt}{4ex}
10  & -0.000\,46            &                      &  297.88                                 & 32.1                                 & 79\,440            &                           & \\
    & -0.000\,438           & 0.229\,6             &  296.910                                & 32.216                               & 79\,924.7          & -2958.67                  &\ 19.601\,4  \\
    & -0.000\,403\,$^{a,c}$ & 0.229\,5\,$^{d,f}$   &  297.623\,$^{c}$                        & 32.620\,$^{c}$                       & 80\,763\,$^{c}$    &      -2976.23\,$^{c}$     &   16.310\,2$^{c}$
\\
%
\rule{-4pt}{5ex}
11  & -0.000\,35            &                      & 399.15                                  & 43.6                                 &  144\,060          &                           &  \\
    & -0.000\,336           &  0.207\,9            & 397.598\                                & 43.68                                &  144\,548          & -4360.21                  &\ 26.409\,5     \\[4pt]
12  & -0.000\,27            &                      & 521.06                                  & 57.6                                 &  247\,579          &                           &   \\
    & -0.000\,264           &  0.189\,9            & 518.810                                 & 57.59                                &  247\,881          & -6208.94                  &\ 34.616\,5     \\[4pt]
13  & -0.000\,21            &                      & 665.54                                  & 74.2                                 &  406\,406          &                           &   \\
    & -0.000\,211           &  0.174\,8            & 662.461                                 & 74.19                                &  406\,633          & -8591.23                  &\ 44.347\,7   \\[4pt]
14  & -0.000\,17            &                      & 834.46                                  & 93.8                                 &  643\,636          &                           &  \\
    & -0.000\,172           &  0.161\,9            & 830.471                                 & 93.72                                &  642\,449          & -11601.2                  &\ 55.726\,6  \\[4pt]
15  & -0.000\,13            &                      & 1\,029.7                                & 116.5                                &  985\,621          &                           &  \\
    & -0.000\,142           &  0.150\,8            & 1\,024.76                               & 116.43                               &  982\,803          & -15340.5                  &\ 68.875\,3   \\[4pt]
16  & -0.000\,11            &                      & 1\,243.7                                & 142.7                                & $1.5  \times 10^6$ &                           & \\
    & -0.000\,12            &  0.141\,1            & 1\,247.27                               & 142.57                               & $1.46 \times 10^6$ & -19918.8                  &\ 83.914\,8 \\[4pt]
17  & -0.000\,09            &                      & 1\,506.9                                & 172.5                                & $2.1  \times 10^6$ &                           & \\
    & -0.000\,099           &  0.132\,6            & 1\,499.91                               & 172.37                               & $2.12 \times 10^6$ & -25453.2                  &\ 100.964  \\[4pt]
18  & -0.000\,08            &                      & 1\,792.7                                & 206.2                                & $3.0  \times 10^6$ &                           &\\
    & -0.000\,084           &  0.125\,0            & 1\,784.65                               & 206.09                               & $3.01 \times 10^6$ & -32068.9                  &\ 120.140 \\[4pt]
19  & -0.000\,06            &                      & 2\,112.3                                & 244.1                                & $4.2  \times 10^6$ &                           &\\
    & -0.000\,070           &  0.118\,2            & 2\,103.40                               & 243.99                               & $4.20 \times 10^6$ & -39898.5                  &\ 141.559  \\[4pt]
20  & -0.000\,05            &                      & 2\,467.7                                & 286.3                                & $5.8  \times 10^6$ &                           &\\
    & -0.000\,062           &  0.112\,1            & 2\,458.12                               & 286.31                               & $5.75 \times 10^6$ & -49082.9                  &\ 165.336 \\
\hline
\end{tabular}}}
\end{center}
\begin{tablenotes}
\item [$\dagger$]{\footnotesize results from \cite{ThakkarSmith:77} and \cite{Pekeris:58} differ from this result in the 4th d.d.}
\end{tablenotes}
\end{threeparttable}
\end{minipage}
\end{table}

\noindent
{\it see (\ref{majorana-2}), with smooth, easily fitted parameters by simple functions in $Z$?}

\begin{figure}
\caption{
\label{fighecusps}
    Cusp parameters {\it vs} nuclear charge $Z$ for He-like systems:\\
   (a) normalized electron-nucleus cusp ${\cal C}_{Z,e}=\frac{C_{Z,e}}{Z}$:  
       from $\Psi_{HK}$ (\ref{HeZ-mod}) with parameters (\ref{parametersHe}), Eq.(\ref{cuspkato:Ze})  (dot-dashed line)
       and from $\Psi_F$ (\ref{psinew}) with parameters (\ref{parametersnew}) (solid line),\\
   (b) electron-electron cusp  $C_{e,e}$: 
       from $\Psi_{HK}$ (\ref{HeZ-mod}) with parameters (\ref{parametersHe}), Eq.(\ref{cuspkato:ee}) (dotted line),
       from $\Psi_F$ (\ref{psinew}) constrained to (\ref{HK:1 to 2})
       (dashed line),
       from $\Psi_F$ (\ref{psinew}) with parameters (\ref{parametersnew})
       \hbox{(solid line)}
}
  \includegraphics[scale=0.4,angle=-90]{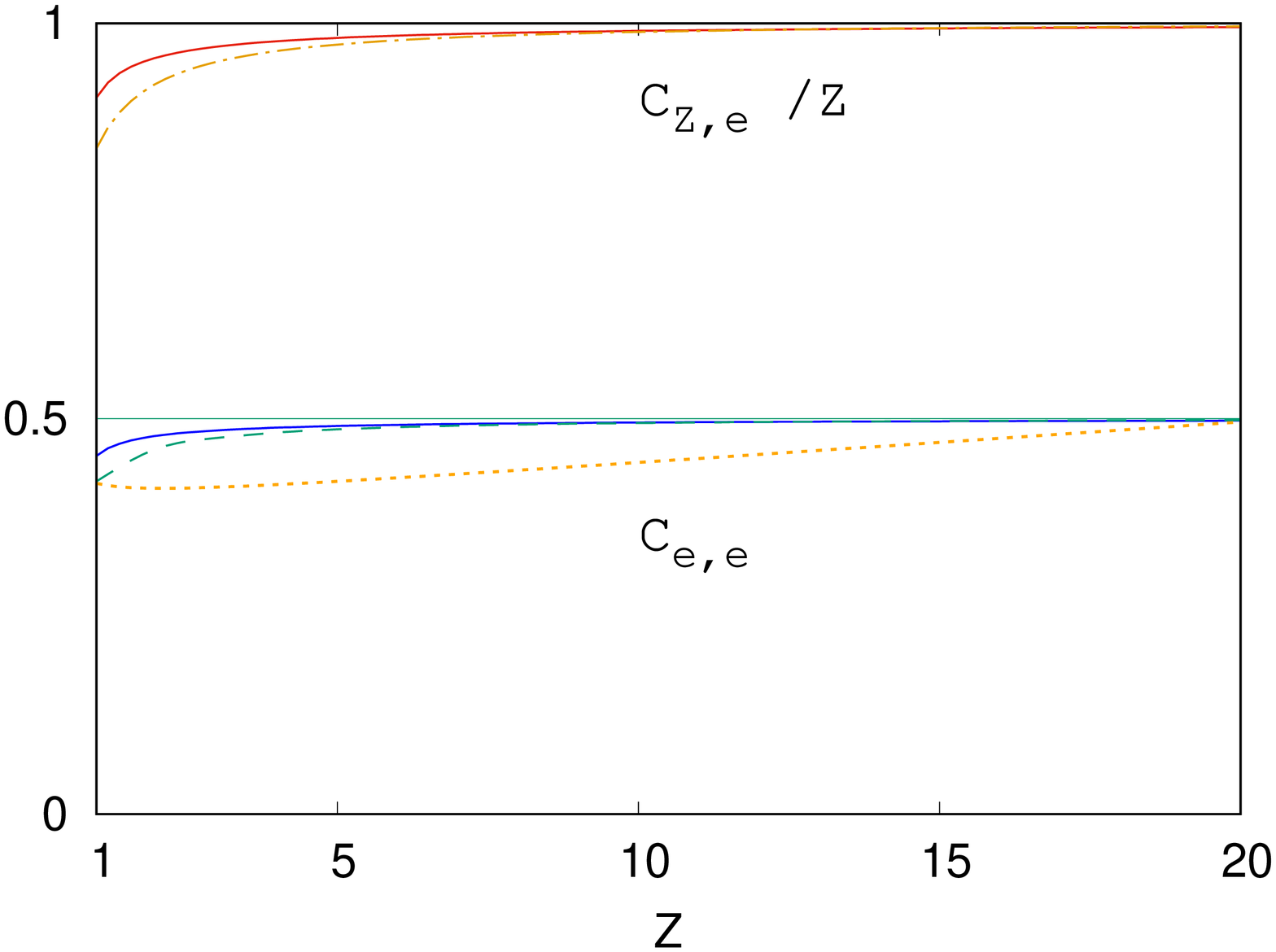}
\end{figure}

Carrying out the constrained variational calculations requiring (\ref{1 to 2}) fulfilled we observe always the smooth behavior of the parameters as functions of the nuclear charge $Z$.
It remains so even if we require a minimal deviation of cusps ${C}^{(F)}_{e,e},{C}^{(F)}_{Z,e}$ for the exact values. The cusps, obtained in this procedure, are fitted by
\begin{align}
C_{e,e}         &\ =\ 1/2\ -\ 0.05/Z\ +\ 0.024/Z^2\ -\ 0.021/Z^3 \ , \non \\
{\cal C}_{Z,e}  &\ = 2 Z C_{e,e} =\ 1\ -\ 0.1/Z\ +\ 0.048/Z^2\ -\ 0.042/Z^3 \ ,
\end{align}
see Fig.\ref{fighecusps}, where ${\cal C}_{Z,e}=C_{Z,e}/Z$ is the normalized electron-nuclear cusp. Fitted parameters for the function (\ref{psinew}) are given by
\begin{align}
\label{parametersnew}
 \al_F     &\ =\ -\, 0.0673/Z\ +\  1.049647\ -\ 0.000842\, Z\ ,  \non \\
 \beta_F   &\ =\ -\, 0.466383/Z\ +\ 0.919016\ +\ 0.001578\,Z\ ,  \non \\
 \gamma_F  &\ =\ -\,\frac{0.704459\ +\ 0.752684\,Z\ -\ 1.52103\,Z^2}
                 {4.278577\ -\ 4.017493\,Z\ -\ 0.388753\,Z^2}\ , \non \\
 a_F       &\ =\ (4 C_{e,e}\ -\ \al_F\ -\ \beta_F)\,Z \ , \non \\
 b_F       &\ =\ C_{e,e}\ -\ \gamma_F \ , \non \\
 c_F       &\ =\  -\,0.036094/Z\ +\ 0.035756\ -\ 0.007766\,Z\ -\ 0.001361\,Z^2 \ ,  \non \\
 d_F       &\ =\  -\,0.090021/Z\ +\ 0.286716\ +\ 0.040912\,Z\ -\ 0.002569\, Z^2 \ ,
\end{align}
cf.(\ref{parametersHe}).

Concrete results for the energies and cusp parameters\ {\it vs}\ $Z$, obtained with the function (\ref{psinew}) and parameters (\ref{parametersnew}), are shown in Table \ref{TABLE-3}: see third rows for each $Z$. We observe that the energies corresponding to the function (\ref{psinew}) are improved (or comparable) with respect to the corresponding values obtained via the function (\ref{HeZ-mod}), see the first rows for chosen $Z$. Absolute deviation from the exact energies, see \cite{AOP:2019} and references therein, increases gradually from $\sim 0.001$\,a.u. for $Z=1$ to $\sim 0.002$\,a.u. at $Z=20$. Simultaneously, the relative deviation from electronic correlation energy $E_{corr}$, see (\ref{Ecorr}), drops gradually from $\sim 0.002$\,a.u. for $Z=1$ to $\sim 0.0002$\,a.u. at $Z=20$.
Both cusps (following the relation (\ref{1 to 2})) are dramatically improved deviating from the exact values in $\sim 10\%$ at $Z=1$ and $\sim 8\%$ at $Z=2$ to $\sim 0.5\%$ for $Z=20$. Making a comparison of the second and third rows for chosen $Z$ in Table \ref{TABLE-3} we have to note that the requirement to have a smooth easily fitted parameters has its cost: at large $Z$, the cusp parameters are slightly worse than the ones in third rows. Needless to say the variational (non-fitted) parameters for the function (\ref{psinew}) lead to cusp results which are comparable (or better) than in the second rows. The behavior of normalized electron-nuclear cusp ${\cal C}_{Z,e}$ and electron-electron cusp $C_{e,e}$ for $Z \in [1, 20]$ are shown on Fig.\ref{fighecusps} for the function $\Psi_{HK}$ (\ref{HeZ-mod}) with parameters (\ref{parametersHe}) and for the function (\ref{psinew}) with parameters (\ref{parametersnew}).

Following our assumption, the {\it maximal} relative local deviation of the function (\ref{psinew}) with parameters (\ref{parametersnew}) from the exact ground state eigenfunction is defined by the relative deviation of the cusps from the exact cusps (residues at Coulomb singularities). It implies that the accuracy of the expectation values is about the same (or better) than the accuracies of the obtained cusp parameters. In Table \ref{expectation1} a number of expectation values are calculated using the function $\Psi_{HK}$ (\ref{HeZ-mod}) with parameters (\ref{parametersHe}) (first rows for each $Z$) and the function (\ref{psinew}) with parameters (\ref{parametersnew}) (second row  for each $Z$) also making the comparison with results from \cite{Harris,Drake,Frolov,ThakkarSmith:77,Pekeris:58,Nakashima:2008}
\footnote{In general, the situation with expectation values is far from being established even for $Z=2$. The list of six papers [5,18-22], in fact, exhausts all studies.
We {\it believe} that the numbers extracted from those papers and presented in Table \ref{expectation1} are correct in all printed digits. Analysis of convergence {\it vs} a number of terms in trial function was performed in \cite{Pekeris:58} {\it only}.  Mysteriously, the results for $<r_{1}>$ presented in \cite{Nakashima:2008}, where benchmark results for energies are obtained, differ systematically from ones in \cite{Drake} by $30-40\%$ for $Z=1,2,3,4$.}.
In general, the above assumption is confirmed! It implies that the function (\ref{psinew}) with parameters (\ref{parametersnew}) can serve as {\it uniform, locally accurate approximation} of the exact ground state eigenfunction. The accuracy in the expectation values can be increased by developing the perturbation theory with respect to the deviation $V_1$ (\ref{PT}).

Using parameters (\ref{parametersnew}) it was also calculated the parameter $A$ in (\ref{cusps-2}),
\[
    A_{F}\ =\ \frac{Z^2}{4}\,(\al_{F}^2+\beta_{F}^2)\ +\ \frac{Z}{2}\, a_{F}\,\al_{F}\ ,
\]
see Table \ref{Table-2}, third rows. A large deviation for $Z=1,2$ from exact values looks surprising.

Note that the {\it same} ultra-compact functions $\Psi_{H,K,HK,F}$ can be used to calculate variational energies for the case when the charge $Z$ is of finite mass, see \cite{twe}.
It will be done elsewhere.

\vspace{-6mm}

\section{Li-like three-electron system: ground state}
\vspace{-5mm}
\subsection{Four trial functions}
\vspace{-5mm}
For Li-like system $(3e; Z)$ we look for a compact function which describes  the ground state of total spin 1/2 in the form
\begin{equation}
\label{GStrialfunct}
 \psi(\vec{r}_1,\vec{r}_2,\vec{r}_3; \chi) = {\cal A} \left[
   \,\phi(\vec{r}_1,\vec{r}_2,\vec{r}_3;\chi) \, \right]\, ,
\end{equation}
where $\chi$ is spin 1/2 eigenfunction made from three 1/2 spins of electrons,
${\cal A}$ is the three-particle antisymmetrizer
\begin{equation}
\label{Asym}
 {\cal A}\ =\ 1 - P_{12} - P_{13} - P_{23} + P_{231}  + P_{312}\, ,
\end{equation}
where $P_{ij}$ represents the permutation $i \leftrightarrow j$, and $P_{ijk}$ stands for the permutation of $(123)$ into $(ijk)$ \footnote{Note that the permutations $P_{231}$ and
$P_{312}$ correspond in {\it standard} notation to $P_{123}$ and $P_{132}$, respectively.}.
Function $\phi$ in (\ref{GStrialfunct}) is called the {\it seed} function and is of the form
\begin{equation}
\label{phi}
   \phi(\vec{r}_1,\vec{r}_2,\vec{r}_3; \chi) \ =\ \phi_1(\vec{r}_1,\vec{r}_2,\vec{r}_3) \chi_1 \ +\ C\,\phi_2(\vec{r}_1,\vec{r}_2,\vec{r}_3) \chi_2\ .
\end{equation}
It is spanned on two orthonormal spin 1/2 eigenfunctions $\chi_{1,2}$, the parameter $C$ is introduced for convenience to measure their relative contribution, see below. Following a similar philosophy for selection of trial function for He-like system, we choose a trial function for Li-like system in the form of a product of six Coulomb orbitals modified by linear in $r$ pre-factors. Hence, the explicitly correlated exponential orbital functions $\phi_{1,2}$ modified by pre-factors are of the form
\begin{equation}
\label{Phi-1}
\phi_1(\vec{r}_1,\vec{r}_2,\vec{r}_3; \al^{(1)}_i, \al^{(1)}_{ij}; a_3)  \ =\
 (1 - a_3 r_3)\
 e^{\scriptstyle -\al^{(1)}_1 Z r_1-\al^{(1)}_2 Z r_2-\al^{(1)}_3 Z r_3} \
   e^{\scriptstyle \al^{(1)}_{12} r_{12} + \al^{(1)}_{13} r_{13} + \al^{(1)}_{23} r_{23}} \ ,
\end{equation}
\begin{equation}
\label{Phi-2}
\phi_2(\vec{r}_1,\vec{r}_2,\vec{r}_3; \al^{(2)}_i, \al^{(2)}_{ij}; a_1)  \ =\
 (1 + a_1 r_{1})\
 e^{\scriptstyle -\al^{(2)}_1 Z r_1-\al^{(2)}_2 Z r_2-\al^{(2)}_3 Z r_3} \
   e^{\scriptstyle \al^{(2)}_{12} r_{12} + \al^{(2)}_{13} r_{13} + \al^{(2)}_{23} r_{23}} \, ,
\end{equation}
where $\al_i^{(p)}, \al_{ij}^{(p)}\, j>i=1,2,3\ , p=1,2$\,, $a_1, a_3$ and $C$ (see (\ref{phi})) are free parameters.

Pre-factors in $\phi_{1,2}$ are chosen \footnote{Electrons are labeled by (1,2,3)} assuming that the electrons $(1,2)$ are in total spin zero state in $\phi_{1}$ while the electrons (2,3) are in total spin zero state in $\phi_{2}$, respectively, {\it i.e.} two electrons $(1,2)$ or $(2,3)$ belong to the closed $1s^2$ shell while the remaining electron is on the next shell.
If pre-factors in $\phi_{1,2}$ are absent, $a_1=a_3=0$ and all non-linear parameters in (\ref{Phi-1}), (\ref{Phi-2}) coincide, $\al^{(1)}=\al^{(2)}$, we arrive at the function which was studied in \cite{TGH2009} for the case $Z=3$.
In general, the function (\ref{GStrialfunct}) is 15-parametric, it contains 3 linear and 12 non-linear variational parameters.  The function (\ref{GStrialfunct}) is anti-symmetrized product of $(1s^2,2s)$ (modified by screening) electronic Coulomb orbitals for one-center problem of charge $Z$ and three electrons with the exponential correlation factors $\sim\exp{(\al_{ij}\, r_{ij})}$.

It is known that there are two linearly independent (normalized) spin $1/2$ functions
of mixed symmetry made from three electronic spins 1/2; in particular, if electrons are labeled by (1,2,3), they have the form
\begin{equation}
\label{chi1}
 \chi_1\ =\  \frac{1}{\sqrt{2}} [ {\boldsymbol\alpha}(1)
   {\boldsymbol\beta}(2)    - {\boldsymbol\beta}(1)
   {\boldsymbol\alpha}(2) ]{\boldsymbol\alpha}(3)\,,
\end{equation}
and
\begin{equation}
\label{chi2}
 \chi_2\ =\  \frac{1}{\sqrt{6}} [  2{\boldsymbol\alpha}(1)
   {\boldsymbol\alpha}(2)  {\boldsymbol\beta}(3)   -
   {\boldsymbol\beta}(1)  {\boldsymbol\alpha}(2)
   {\boldsymbol\alpha}(3)  - {\boldsymbol\alpha}(1)
   {\boldsymbol\beta}(2) {\boldsymbol\alpha}(3) ] \,,
\end{equation}
here ${\boldsymbol\alpha}(i)$, ${\boldsymbol\beta}(i)$ are spin up, spin down eigenfunctions of $i$-th electron, respectively. These functions correspond to the Young tableau
\ytableausetup{centertableaux}
{\tiny \begin{ytableau}
1 & 3  \\
2
\end{ytableau}
} and
{\tiny \begin{ytableau}
1 & 2  \\
3
\end{ytableau}
}
respectively. After applying the antisymmetrizer (\ref{Asym}) to the function (\ref{phi})
(see \cite{Pauncz} for details) the result can be written as
\begin{equation}
\label{completetrial}
 \psi\ =\ \left(\hat{P}_{\phi_1\phi_1}\phi_1\ +\ C  \hat{P}_{\phi_1\phi_2}\phi_2\right)\text{$\chi$}_1\ +\ \left(\hat{P}_{\phi_2\phi_1}\phi_1\ +\  C \hat{P}_{\phi_2\phi_2}\phi_2\right)\text{$\chi$}_2\ ,
\end{equation}
where projectors are defined as following
\begin{eqnarray}
\label{projectors}
\hat{P}_{\phi_1\phi_1}&\ =\ &\frac{1}{2\sqrt{3}}\,[2{I}+2{P}_{12}-{P}_{13}-{P}_{23}-{P}_{231}-{P}_{312}]\ ,\non \\
\hat{P}_{\phi_2\phi_2}  &\ =\ &\frac{1}{2\sqrt{3}}\,[2{I}-2{P}_{12}+{P}_{13}+{P}_{23}-{P}_{231}-{P}_{312}]\ ,\non
\end{eqnarray}
\begin{equation}
\hat{P}_{\phi_2\phi_1}  \ =\ \frac{1}{2}\,[{P}_{13}-{P}_{23}-{P}_{231}+{P}_{312}]\qquad ,\qquad
\hat{P}_{\phi_1\phi_2}  \ =\ \frac{1}{2}\,[{P}_{13}-{P}_{23}+{P}_{231}-{P}_{312}]\ ,
\end{equation}
where $\hat{I}$ is identity operator.
Let us emphasize that even if one $\phi$'s vanishes, $\phi_1 (\phi_2)=0$, after the antisymmetrization both spin function contributions $\chi_{1,2}$ emerge in the resulting function (\ref{completetrial}).
Projectors $\hat{P}_{\phi_1\phi_1}$ and $\hat{P}_{\phi_2\phi_2}$ agree with those presented explicitly in \cite{Frolov:2010}. However, $\hat{P}_{\phi_1\phi_2}$ and $\hat{P}_{\phi_2\phi_1}$
are {\it not} equal as was claimed in \cite{Frolov:2010}.
%
It is worth mentioning a particular case of factorizable orbitals, corresponding to the physics picture of shell model: a closed shell plus a valence electron orbital i.e.
$$\phi_2(\vec{r}_1,\vec{r}_2,\vec{r}_3)=
\varphi_{0}(\vec{r}_1)\varphi_{0}(\vec{r}_2)\varphi_{1}(\vec{r}_3)\ ,$$
where terms $\hat{P}_{\phi_1\phi_2}\phi_2$ and $\hat{P}_{\phi_2\phi_2}\phi_2$ vanish and therefore only $\phi_1$ contributes to the energy.

Besides the general case we also study two important particular cases of the ground state function (\ref{phi})-(\ref{Phi-2}). They occur when some constraints on the parameters are imposed:
\begin{enumerate}
[label=\alph*]
\item[{\bf (a)}]   $a_1=a_3=0$,  $\al_{i}^{(1)}=\al_{i}^{(2)} \equiv \al_i$,
                 	$\al_{ij}^{(1)}=\al_{ij}^{(2)}\equiv \al_{ij}$, $i=1,2,3$, $j<i$\,.\\
                    In this case the orbital parts are equal, $\phi_1=\phi_2 \equiv \varphi$ and the trial function appears in a factorized form as orbital function multiplied by a linear superposition of spin $1/2$ functions,
\begin{equation}
    \label{phichi1}
	 \phi(\vec{r}_1,\vec{r}_2,\vec{r}_3)\chi= \varphi(\vec{r}_1,\vec{r}_2,\vec{r}_3; \al_1,
     \al_2,\ldots)\,( \chi_1 + C\chi_2 )\ ,
\end{equation}
where
\begin{equation}
\label{varphichi1}
	 \varphi(\vec{r}_1,\vec{r}_2,\vec{r}_3; \al_1, \al_2,\ldots) =
	  e^{\scriptstyle -\al_1 Z r_1-\al_2 Z r_2-\al_3 Z r_3} \
      e^{\scriptstyle \al_{12} r_{12} + \al_{13} r_{13} + \al_{23} r_{23}} \ .
\end{equation}
At $Z=3$ it coincides to the trial function $\psi_1$ which was used for the first time in \cite{TGH2009} to study the lithium atom~
\footnote{Note that in \cite{TGH2009} the value of the variational energy was printed $-7.4547$, which is slightly different from the present one $-7.4544$ in 4th d.d. (see Table~\ref{results-Li}). It occurred due to a loss of accuracy in six-dimensional integration: in present work it was verified by making calculations with higher accuracy.}. In total, this function contains 7 variational parameters, $\al_1,\al_2,\ldots$ and $C$, which {\it measures} a relative contribution of the spin $1/2$ eigenfunction $\chi_2$ in (\ref{phichi1}), see also (\ref{phi}).
Recently, this trial function was used to study the Li-like sequence for $Z \in [3,14]$  \cite{AOP:2019} in a search for a compact function which would be able to describe the domain of applicability of the QMCC in static approximation. In that study it was observed that the weight parameter $C$ in  (\ref{phichi1}) grows monotonically with $Z$ from $\sim 0.05$ at $Z=3$ to $\sim 0.5$ for $Z=14$. This result suggests the importance of the second spin function $\chi_2$ contribution in (\ref{phichi1}), which we will use in subsequent trial functions {\bf (b)} and {\bf (c)}. In present study we have repeated the calculations by demanding a more smooth behavior of the variational parameters as a function of $Z$ which surprisingly improved the results of \cite{AOP:2019} in 4-5th figures (see Table~\ref{results-Li}).
Naturally, the function (\ref{phichi1}) provides a rough approximation for the parameters for more general cases {\bf (b)} and {\bf (c)} (see below).

\item[{\bf (b)}]  $a_1=a_3=0$\,.\\
      In this case the non-anti-symmetrized, {\it seed} function (\ref{phi})  contains the same type of orbital function $\varphi(\vec{r}_1,\vec{r}_2,\vec{r}_3; \al_1, \al_2,\ldots)$, as in (\ref{varphichi1}), multiplying each spin eigenfunction $\chi_{1,2}$. Each orbital function is characterized by their own variational parameters,
      {\it i.e.} \hbox{$\phi_1=\varphi(\vec{r}_1,\vec{r}_2,\vec{r}_3; \al_1^{(1)}, \al_2^{(1)},\ldots)$} and \hbox{$\phi_2=\varphi(\vec{r}_1,\vec{r}_2,\vec{r}_3; \al_1^{(2)}, \al_2^{(2)},\ldots)$}. These orbital functions are products of (screened) $1s$ Coulomb orbitals $\sim \exp{(-\al_i r_i)}$ and explicitly correlated exponential functions $\sim \exp(\al_{ij}r_{ij})$, $\scriptstyle j>i=1,2,3$. The Pauli principle is guaranteed since each $1s$-{\it orbital} has different parameters (the same argument applies to the trial function {\bf (a)} given by the function (\ref{phichi1})). After anti-symmetrization, the trial function (\ref{completetrial}) contains both spin functions $\chi_{1,2}$ each multiplied by certain combinations of spatial orbitals $\phi_{1,2}$ of mixed symmetry. In total, we have 13 variational parameters, including the parameter $C$. The (non-symmetrized) seed trial function takes the form
\begin{equation}
\label{phichi2}
	   \phi(\vec{r}_1,\vec{r}_2,\vec{r}_3)\chi\ =\
	   \varphi(\vec{r}_1,\vec{r}_2,\vec{r}_3; \al_1^{(1)}, \al_2^{(1)},\ldots) \chi_1\ +\
	   C \varphi(\vec{r}_1,\vec{r}_2,\vec{r}_3; \al_1^{(2)}, \al_2^{(2)},\ldots)\chi_2 \ ,
\end{equation}
	 cf. (\ref{phichi1}), where the functions $\varphi$'s in r.h.s. are of the form
     given in (\ref{varphichi1}).

\item[{\bf (c)}]  In the general case the orbital functions $\phi_{1,2}$ in (\ref{phi})
      correspond to the product of two $(1s) \sim \exp(-\al r)$ and one
      $(2s)\sim $ \hbox{$(1+a r)\exp(-\al r)$} modified by screened Coulomb orbitals, multiplied by (explicitly correlated) exponential functions $\sim \exp(\al_{ij}r_{ij})$, which mimics a Coulomb repulsion of electrons. Thus, two electrons occur in spin-singlet state forming a closed shell with approximately equal distances to the heavy center while 3rd electron is situated far away. In particular, the function $\phi_1$ describes the situation when the far-distant electron labeled by 3 is given by the $(2s)$ orbital while as for the function $\phi_2$ the electron labeled by 1 belongs the $(2s)$ orbital. In total we have 15 variational  parameters including the parameter $C$. Eventually, the seed (non-symmetrized) function $\phi$ is of the form (\ref{phi}) with $\phi_1,\phi_2$ given by (\ref{Phi-1}) and (\ref{Phi-2}), respectively.

\end{enumerate}

By using three properties of the antisymmetrizer:
(i) $\hat{A}$ commutes with the Hamiltonian ${\hat H}$,
(ii) Hermiticity $(\hat{A}=\hat{A}^{\dagger})$ and
(iii) idempotency $(\hat{A}^2=\hat{A})$,
the variational energy $E_{var}$ can be written as the ratio of two nine-dimensional integrals
\begin{equation}
\label{expectationfinal}
 E_{var}\ =\ \frac{ \int[\phi_1^{*}\hat{H}\hat{P}_{\phi_1\phi_1}
 \phi_1 +
 \phi_1^{*}\hat{H}\hat{P}_{\phi_1\phi_2}\phi_2+\phi_2^{*}\hat{H}\hat{P}_{\phi_2\phi_1}\phi_1+
\phi_2^{*}\hat{H}\hat{P}_{\phi_2\phi_2}\phi_2]\,dV}{\int[\phi_1^{*}\hat{P}_{\phi_1\phi_1}\phi_1+
  \phi_1^{*}\hat{P}_{\phi_1\phi_2}\phi_2+\phi_2^{*}\hat{P}_{\phi_2\phi_1}\phi_1+
  \phi_2^{*}\hat{P}_{\phi_2\phi_2}\phi_2]\,dV}\ ,
\end{equation}
where $dV$ is volume element.
Note that in (\ref{expectationfinal}) the parameter $C$, see (\ref{phi}), (\ref{completetrial}), has been absorbed into $\phi_2$.

After separation of the center-of-mass, in the space of relative coordinates of the general four-body problem $(r_1, r_2, r_3, r_{12}, r_{13}, r_{23}, \Om(\tha_1, \tha_2, \tha_3))$,
the integration over three angles in $\{\Om\}$, describing overall orientation and rotation of the system, are easily performed analytically. Hence, the nine-dimensional integrals in (\ref{expectationfinal}) can be also reduced to six-dimensional integrals over the relative distances $(r_1, r_2, r_3, r_{12}, r_{13}, r_{23})$ (for the general discussion see \cite{twe}). It was shown a long ago \cite{Fromm-Hill} that these integrals can be reduced to a combination of one-dimensional ones (!) but with integrands involving dilogarithm functions.
The analytic properties of the resulting expressions for these integrands are found to
be unreasonably complicated (see e.g. \cite{Harris}) for numerical evaluation.
For that reason we used the direct numerical evaluation of the six-dimensional integrals, by means of
a numerical adaptive multidimensional integration routine (Cubature). As for the minimization we employ the routine MINUIT from CERN-LIB. A dedicated code was written in \verb!C++! with MPI paralellization. The program was executed in the cluster KAREN at ICN-UNAM (Mexico) using 120 Xeon processors with working frequencies 2.70\,GHz each.

It has to be noted that it was found recently that the ground state energy of the Li-like systems as a function of $Z$ in domain $Z \in [3, 20]$ follows accurately the Majorana-type formula \cite{AOP:2019},
\begin{equation}
\label{majorana-3}
 E^{(3)}_{M}(Z)\ =\ -\frac{9}{8} Z^2 + 1.02326 Z -0.416432\ ,
\end{equation}
in a.u., cf. (\ref{majorana-2}). This fit reproduces at least 3-4 s.d. in the ground
state energy leading, in fact, to the exact QMCC energies at $Z \in [3, 20]$.
In particular, for Lithium atom $Z=3$ the Majorana formula gives $E^{(3)}_{M}(3)\,=\, -7.472$\,a.u. {\it vs} $-7.478$\,a.u. as for the exact result, while for $Z=10$  $E^{(3)}_{M}(10)\,=\,-102.6838$\,a.u. {\it vs} $-102.6822$\,a.u. as for the exact result
(see Ref. \cite{AOP:2019} for discussion).

Table \ref{results-Li} shows the results of variational calculations of the ground state for
Li-like atoms obtained with the trial functions {\bf (a)} (\ref{phichi1}),  {\bf (b)} (\ref{phichi2}) and  {\bf (c)} (\ref{phi})-(\ref{Phi-2}), see 3rd column of Table, compared with the accurate results known in literature, see 2nd column. The results demonstrate that the simplest trial function  {\bf (a)} describes 2 s.d. of the ground state energy for $Z=3,4$ and it improves gradually with increasing $Z$ up to 3-4 s.d. As for the function (b) the accuracy is (slightly) increased to $\sim 3-4$ s.d. (without rounding) uniformly in the whole range  $Z=3\ldots 20$. In turn, the function {\bf (c)} increases essentially the accuracy
of the variational energy with respect to the {\it exact} value. For Lithium atom the relative difference is reduced from $\sim 0.32\%$ as for the function {\bf (a)} to $\sim 0.06\%$, as for ${\rm  Ne}^{7+}$ ($Z=10$) from $\sim 0.42\%$ of the function {\bf (a)}, up to $\sim 0.007\%$
(see below). Thus, the more general compact function {\bf (c)} allow to reproduce at least 4 s.d. of the exact energy in the entire range $Z=3-20\,$ i.e. of the domain of applicability of QMCC. The variational parameters behave smooth as a function of the nuclear charge $Z$:
they can be fitted accurately by a second degree polynomials in $Z$, see Table \ref{params-Li}
(for {\bf (a)} and {\bf (c)}).

In order to check the local accuracy of the compact trial functions (\ref{phichi1}),   (\ref{phichi2}), (\ref{phi})-(\ref{Phi-2}), we evaluate the nuclear-electron $C_{Z,e}$ and  electron-electron $C_{e,e}$ cusp parameters defined as follows
\begin{equation}
\label{cusp1}
C_{Z,e}\ \equiv\ -\ \frac{\langle \delta({\bf r}_i)\frac{\pa}{\pa r_i} \rangle}{\langle
\de({\bf r}_i) \rangle}\ , \qquad i= 1,2,3\ ,
\end{equation}
see e.g. \cite{TGH2009}, and
\begin{equation}
\label{cusp2}
C_{e,e}\ \equiv\ \frac{\langle \de({\bf r}_{ij})\frac{\pa}{\pa r_{ij}} \rangle}{\langle
\de({\bf r}_{ij}) \rangle}\ , \qquad j>i=1,2,3 \ .
\end{equation}
Based on the Schr\"odinger equation these parameters should take values $Z$ and $1/2$, respectively. The values of the cusp parameters derived from the variational trial functions (\ref{phichi1}), (\ref{phichi2}), (\ref{phi})-(\ref{Phi-2}) are shown in Table \ref{results-Li}.
These results indicate that trial functions {\bf (a), (b) \rm and \bf (c)} with parameters found variationally describe with accuracy $\lesssim 1-2\%$ the nuclear-electron cusp ${\cal C}_{Z,e}$ in domain $Z \in [3, 20]$, but fail to describe adequately the electron-electron cusp ${\cal C}_{e,e}$ - the deviation from the exact value is $\sim 40 - 50\%$\
(see Table~\ref{results-Li}). Similarly to two-electron case, the fact that the electron-electron cusp at $r_{ij}=0$ is not well-reproduced by the functions {\bf (a), (b) \rm and \bf (c)} implies that they do not behave correctly in vicinity of $r_{ij}=0$. However, it does not
influence the quality of the variational energy. One can see that the improvement of variational energy not necessarily leads to improvement in a description of the cusp parameters.
It turns out that the function {\bf (b)} provides systematically the most accurate values of cusp parameters in comparison with {\bf (a) \rm and \bf (c)}. Meanwhile, the function {\bf (c)}
gives the most accurate variational energies.

We were unable to find a modification of parameters of the function {\bf (c)}  (\ref{phi})-(\ref{Phi-2}), see Table \ref{params-Li}, which would allow us to improve the accuracy in electron-electron cusp $C_{e,e}$ without loosing accuracy in energy $E_{exp}$ and electron-nuclear cusp $C_{Z,e}$. It definitely requires a further generalization of the function {\bf (c)} by adding linear in $r_{ij}$ terms into pre-factors in (\ref{Phi-1})-(\ref{Phi-2}) and/or modifying some inter-electron $r_{ij}$-terms in Coulomb orbital's exponents. Evidently, such a modification should support their linear in $r$ behavior at small and large $r$. The simplest modification, which fulfils this requirement, is a rational function,
\begin{equation}
\label{3e-modification}
    \al_{ij}\, r_{ij} \rar \al_{ij}\, r_{ij}\ \frac{1\ +\ c_{ij}\, r_{ij}}{1\ +\ d_{ij}\, r_{ij}}\ \equiv\ \al_{ij}\, {\hat r}_{ij} \ ,
\end{equation}
cf.(\ref{He-rational}) for the case of helium sequence, as was proposed in \cite{BMBM:2001},
where $\{\al, c, d\}$ are parameters.
\begin{enumerate}
\item[{\bf (d)}]\ Following above-mentioned physics arguments a minimal modification of function {\bf (c)} should be
\[
 \phi_1(\vec{r}_1,\vec{r}_2,\vec{r}_3;\ \al^{(1)}_i, \al^{(1)}_{ij}; a_3,\  b^{(1)}_{12}, c_{12}, d_{12})  \ =\
\]
\begin{equation}
\label{Phi-1-mod}
 (1 - a_3 r_3 + b^{(1)}_{12}\, r_{12})\
   e^{-\al^{(1)}_1 Z r_1-\al^{(1)}_2 Z r_2-\al^{(1)}_3 Z r_3} \
   e^{\al^{(1)}_{12} {\hat r}_{12} + \al^{(1)}_{13} r_{13} + \al^{(1)}_{23} r_{23}} \ ,
\end{equation}
as for the first orbital function, hence, the inter-electron interaction between (1,2) electrons
in spin-singlet state is modified by introducing the term $b^{(1)}_{12}\, r_{12}$ into pre-factor and $r_{12}$-dependent screening in rational form, see (\ref{3e-modification}), in exponent, and
\[
 \phi_2(\vec{r}_1,\vec{r}_2,\vec{r}_3;\ \al^{(2)}_i, \al^{(2)}_{ij}; a_1,\  b^{(2)}_{23}, c_{23}, d_{23})  \ =\
\]
\begin{equation}
\label{Phi-2-mod}
 (1 + a_1 r_{1} + b^{(2)}_{23}\, r_{23})\
 e^{\scriptstyle -\al^{(2)}_1 Z r_1-\al^{(2)}_2 Z r_2-\al^{(2)}_3 Z r_3} \
   e^{\scriptstyle \al^{(2)}_{12} r_{12} + \al^{(2)}_{13} r_{13} + \al^{(2)}_{23} {\hat r}_{23}} \, ,
\end{equation}
as for the second orbital function, hence, the inter-electron interaction between (2,3) electrons in spin-singlet state is modified by introducing the term $b^{(2)}_{23}\, r_{23}$ into pre-factor and $r_{23}$-dependent screening in rational form, see (\ref{3e-modification}), in exponent in the seed function (\ref{phi}),
\vspace{-3mm}
%
%
\[
   \phi^{\bf (d)}(\vec{r}_1,\vec{r}_2,\vec{r}_3; \chi) \ =\ \phi_1(\vec{r}_1,\vec{r}_2,\vec{r}_3;\ \al^{(1)}_i, \al^{(1)}_{ij}; a_3,\  b^{(1)}_{12}, c_{12}, d_{12})\, \chi_1 \ +
\]
\begin{equation}
\label{Phi-d}
 C\,\phi_2(\vec{r}_1,\vec{r}_2,\vec{r}_3;\ \al^{(2)}_i, \al^{(2)}_{ij}; a_1,\  b^{(2)}_{23}, c_{23}, d_{23})\, \chi_2\ .
\end{equation}
Here $\al_i^{(p)}, \al_{ij}^{(p)},\ j>i=1,2,3\ , p=1,2$\ , and
$a_1, a_3$, $b^{(1)}_{12}, b^{(2)}_{23}$,
$c_{12}, d_{12}, c_{23}, d_{23}$  and the weight factor $C$ are free parameters.
In total, it is characterized by 21 free parameters. If the parameters $$a_1=a_3=b^{(1)}_{12}=b^{(2)}_{23}=c_{12}=d_{12}=c_{23}=d_{23}=0\ ,$$
the Anzatz {\bf (d)} (\ref{Phi-d}) degenerates to the Anzatz {\bf (b)}.
If the parameters $$b^{(1)}_{12}=b^{(2)}_{23}=c_{12}=d_{12}=c_{23}=d_{23}=0\ ,$$
the Anzatz {\bf (d)} (\ref{Phi-d}) degenerates to the Anzatz {\bf (c)}.
\end{enumerate}

\vspace{-9mm}

\subsection{Results}

\vspace{-5mm}

\begin{figure}[tb!]
\begin{center}
      \includegraphics[angle=-90,scale=0.4]{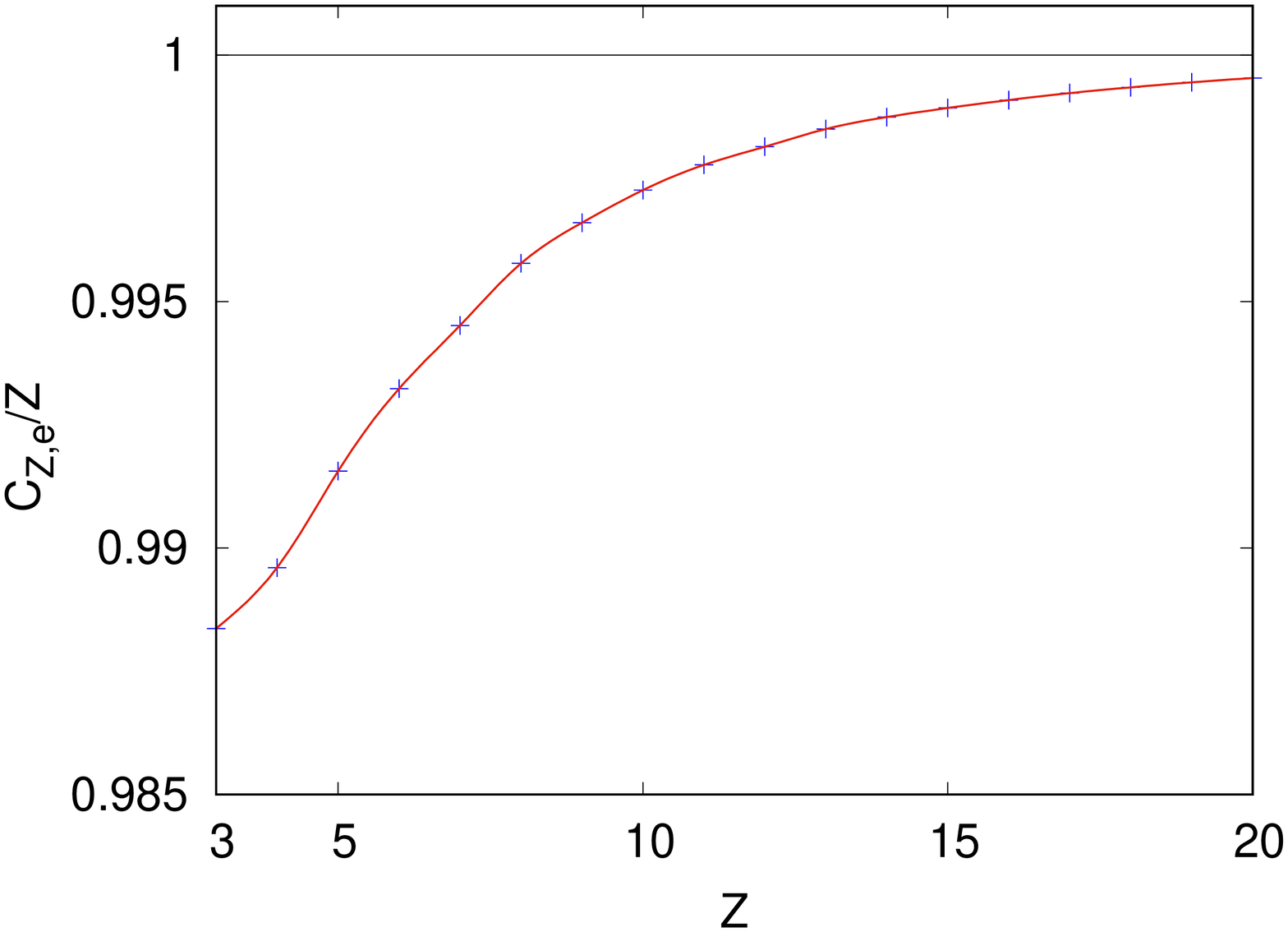}
      \caption{Lithium-like atomic ions: normalized electron-nuclear cusp (marked by crosses) ${\cal C}_{Z,e}=C_{Z,e}/Z$ {\it vs} $Z$ for function ${\bf (d)}$. Curve is the interpolation by splines.}
\label{Li-cuspne}
\end{center}
\end{figure}

\begin{figure}[thb!]
\begin{center}
      \includegraphics[angle=-90,scale=0.4]{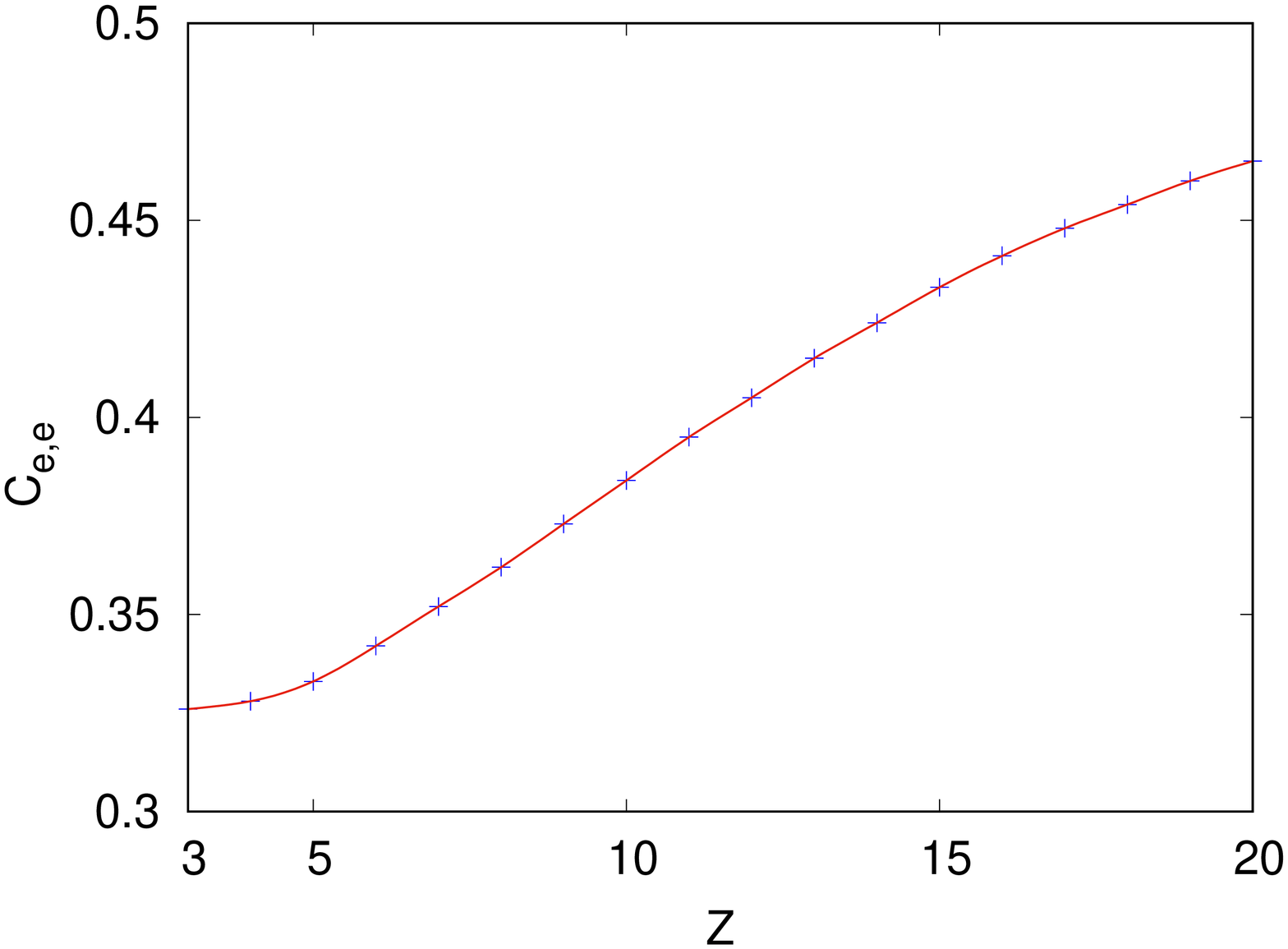}
\caption{Lithium-like atomic ions: electron-electron cusp $C_{e,e}$ (marked by crosses) {\it vs} $Z$ for function ${\bf (d)}$. Curve is the interpolation by splines.
}
\label{Li-cuspee}
\end{center}
\end{figure}

The results of variational calculations with function {\bf (d)} (\ref{Phi-d}) are shown
in Table \ref{results-Li}. In the whole domain $Z \in [3, 20]$ the deviation of the variational energy $E_{var}$ from the exact one occurs in the 3rd d.d. being much more accurate than the results {\bf (a), (b), (c)}. It fully reproduces the domain of applicability of QMCC - the domain, which is free of any corrections (and beyond)!
At $Z=3$ the variational energy differs from the exact one, see e.g. \cite{AOP:2019}, in $\sim 0.03\%$ and with growth of $Z$ it reduces quickly to $\sim 0.002\%$ at $Z=20$. It is demonstrated that the electronic correlation energy - the difference between total energy and the sum of the energies of three hydrogen atoms where two are in its ground state and third one is in the first excited state -
\begin{equation}
\label{Ecorr-3}
  E_{corr}\ =\ E\ +\ \frac{9\,Z^2}{8}\ (\approx\ -\,0.153282\ +\ 0.624583\, Z)\ ,
\end{equation}
see (\ref{majorana-3}), grows linearly with $Z$, while it is relative deviation
\[
   \frac{\De E}{E_{corr}}\ =\ \frac{E_{var}-E_{exact}}{E_{corr}}\ ,
\]
is of the order of $10^{-4}$ for $Z \leq 20$. It decreases with growth of $Z$.
Electron-nuclear cusp $C_{Z,e}$, see Fig.\ref{Li-cuspne}, differs from exact one in $\sim 1\%$ at $Z=3$ and then gradually reduces to $\lesssim 0.1\%$ at $Z=20$. In turn, the electron-electron cusp, see Fig.\ref{Li-cuspee}, differs in $\sim 35\%$ for $Z=3$; then this difference reduces gradually to $\sim 3\%$ for $Z=20$. A striking fact is that all variational parameters behave smoothly and monotonously {\it vs}\ $Z$: they are easily interpolated by 2nd degree polynomials in $Z$, except for $C$, which is interpolated by the 2nd degree polynomial in $1/Z^2$, see Table \ref{params-Li}. In Table \ref{exp-val-Li} the expectation values for inter-electron distances, $\langle r_{ij}\rangle $ are $\langle r^{-1}_{ij}\rangle $, are presented. Comparison with results by Yan-Drake \cite{Yan-Drake:1995} available for $Z=3$ shows the coincidence in 3 s.d., the difference occurs in the 4th s.d.

Now let us make the general remark about significance of our choice for the trial function (\ref{phi}). It has to be mentioned there were numerous discussions whether the second spin
function $\chi_2$ (or, equivalently, $\phi_2 \neq 0$) should be included in the general trial
function (see, for instance, \cite{Larsson}).
After all, the full anti-symmetrization of the trial function (\ref{GStrialfunct})-(\ref{phi})
takes care of this fact and generates the total spin $1/2$ function (\ref{completetrial}),
where both spin functions $\chi_{1,2}$ are present and furthermore they appear with equal weights (!), see below. Deviation from the equality indicates the accuracy the accuracy of the calculation
of the  variational integrals. The probability of finding each spin function can be computed as
\begin{equation}
 \label{probspin}
 P_{{\chi_{1,2}}}\ =\ \int|\langle \chi_{1,2}|\psi\rangle|^2\,dV \quad ,
 \quad P_{{\chi_{1}}}+P_{{\chi_{2}}}=1   \ .
\end{equation}
The probabilities (\ref{probspin}) for the trial functions {\bf (a), (b), (c) \rm and \bf (d)} are
calculated and are shown in Table \ref{results-Li}. It can be noticed  that the probability of finding spin functions $\chi_{1,2}$ are very close to $\sim 1/2$ each independently
on the trial function used (case {\bf (a), (b), (c) \rm or \bf (d)}). Nonetheless,
the most general function (case {\bf (d)}) fulfills this condition almost exactly.
In general, it seems that $|\langle \chi_2|\psi\rangle|^2$  and $|\langle \chi_1|\psi\rangle|^2$ are related via a certain permutation, which leaves the volume element $dV$ invariant. Hence, there must exist an operator $\hat{C}$ such that
\begin{equation}
\label{conditionC}
     \hat{C}(\hat{P}_{\phi_1\phi_1}\phi_1\ +\ \hat{P}_{\phi_1\phi_2}\phi_2)^2\ =\ (\hat{P}_{\phi_2\phi_1}\phi_1\ +\ \hat{P}_{\phi_2\phi_2}\phi_2)^2\ ,
\end{equation}
which can be found explicitly,
\begin{equation}
\label{C}
\hat{C}\ =\ -{I}+\frac{2}{3}({P}_{12}+{P}_{13}+{P}_{23})\ .
\end{equation}
It explains why the equality
\[
  P_{{\chi_{1}}}\ =\ P_{{\chi_{2}}}\ =\ 1/2\ ,
\]
should hold. Deviations from this equality seen in Table \ref{results-Li} are consequences of inaccuracies in multidimensional integration.

\vspace{-1.6mm}

By attempting to evaluate the contribution of $\phi_2$ from the seed function (\ref{phi}) to the final result we have calculated a relative contribution to the normalization of the original trial function with $\phi_2 \neq 0$ coming from the function when $\phi_2 = 0\ (C=0)$,
\begin{equation}
\label{deltaN}
 \De N\ =\ \Bigg\vert{ 1 \ -\ \frac{N_{(C=0)}}{N}}\Bigg\vert \ ,
\end{equation}
where $N=\int|\psi|^2\,dV\,$ and $N_{(C=0)}=\int|\psi(C=0)|^2\,dV\,$. The dependence on the relative contribution to the normalization $\De N$ for different trial functions
{\bf (a), (b), (c), (d)}, see Table \ref{results-Li}, is checked. Systematically, the largest $\De N$ corresponds to case {\bf (b)} for $Z \in [3,10]$, while for the case {\bf (c)} the smallest $\De N$ occurs at $Z > 6$ being, in general, of the order $10^{-1}$ for $Z < 10$ and dropping to $\sim 10^{-2}$ for $Z > 10$. As for the most accurate function {\bf (d)} the relative contribution $\De N$ to the normalization integral is {\it always} of the order of $\sim 10^{-2}$.

\vspace{-1.8mm}

It has to be emphasized that the {\it same} ultra-compact functions {\bf (a), (b), (c), (d)}, but with different parameters, can be used to calculate variational energies for the Li-like 3-electron sequence even for the case of finite mass of the charge $Z$, see \cite{twe}.
It will be done elsewhere.

\begin{table}[]
\caption{\footnotesize Ground state $1\,{}^20^+$ of Li-like atomic systems: \\[6pt]
 	    Exact energy $E_{ex}$, extracted from \cite{Puchalski}, compared with variational
        energies $E_{v}^{(\rm a, b, c, d)}$ for different compact functions (see text);
        correlation energy $E_{corr}$ (see text) and absolute deviation $\De E/E_{corr}=(E_{v}-E_{ex})/E_{corr}$ shown;
 	    normalized electron-nuclear cusp parameter $\mathcal{C}_{Z,e} \equiv C_{Z,e}/Z$
        ($\mathcal{C}^{(exact)}_{Z,e}=1$) and electronic cusp parameter $C_{e,e}$
         ($C^{(exact)}_{e,e}=1/2$) presented;\\
 	    probabilities $P_{\chi_1}$ and difference in the normalization $\De N$ (see text) shown.}
\label{results-Li}
\begin{center}
{\setlength{\tabcolsep}{0.6cm} \renewcommand{\arraystretch}{1}
 \resizebox{0.95\textwidth}{!}{%
 		\begin{tabular}{ccccccccc}
 			\hline
 			$\quad Z$ & $E_{ex}$(a.u.) & $E_v$(a.u.) & $E_{corr}$(a.u.) & $\De E/E_{corr}$ &
             $\mathcal{C}_{Z,e} (=1)$ & $C_{e,e} (=1/2)$ & $P_{\chi_1}$ & $\De N$\\
\hline
 			\rule{0pt}{4ex}3 & -7.4781 &-7.4544$^\text{a}$& & &0.984$^\text{a}$ & 0.216$^\text{a}$&0.5151$^\text{a}$&0.1687$^\text{a}$\\
 			&& -7.4700$^\text{b}$& &&0.997$^{\text{b}}$&0.222$^{\text{b}}$
 			&0.5144$^\text{b}$&
 			0.2298$^\text{b}$\\
 			&&-7.4733$^{\text{c}}$& &&0.988$^{\text{c}}$&0.221$^{\text{c}}$
 			&0.4996$^\text{c}$ & 0.0340$^\text{c}$ \\
 			&&-7.4757$^{\text{d}}$& 2.6493$^\text{d}$&0.00091$^{\text{d}}$&0.988$^{\text{d}}$&0.326$^{\text{d}}$
 			&0.5001$^\text{d}$&0.0546$^\text{d}$\\[5pt]
\hline
 			\rule{0pt}{5ex} 4& -14.3248 &-14.2718$^\text{a}$& & & 0.982$^\text{a}$& 0.222$^\text{a}$&0.5181$^\text{a}$&0.0083$^\text{a}$ \\
 			& &-14.3059$\text{b}$& & & 0.995$^\text{b}$&
 			0.230$^\text{b}$&
 			0.5168$^\text{b}$&
 			0.0861$^\text{b}$\\
 			&&-14.3194$^{\text{c}}$& & &0.989$^{\text{c}}$&0.217$^{\text{c}}$
 			&0.5017$^\text{c}$&0.0431$^\text{c}$\\
 			&&-14.3219$^{\text{d}}$& 3.6781$^\text{d}$&0.00079$^{\text{d}}$ & 0.990$^{\text{d}}$&0.328$^{\text{d}}$
 			&0.5001$^\text{d}$ &0.0409$^\text{d}$\\[5pt]
\hline
 			\rule{0pt}{5ex} 5&-23.4246  & -23.3336$^\text{a}$ & & &0.981$^\text{a}$&0.227$^\text{a}$
 			&0.5108$^\text{a}$&
 			0.0127$^\text{a}$\\
 			&&-23.3931$^\text{b}$& && 0.994$^\text{b}$& 0.241$^\text{b}$&
 			0.5116$^\text{b}$&
 			0.0872$^\text{b}$\\
 			& &-23.4186$^{c}$& && 0.990$^{\text{c}}$&0.220$^\text{c}$&0.4997$^\text{c}$
 			& 0.0335$^\text{c}$\\
 			&&-23.4216$^{\text{d}}$& 4.7034$^\text{d}$ & 0.00064$^\text{d}$&0.992$^{\text{d}}$&0.333$^\text{d}$&0.5001$^\text{d}$
 			& 0.0245$^\text{d}$
 			\\[5pt]
 			\hline
 			\rule{0pt}{5ex} 6& -34.7755 &-34.6381$^\text{a}$ && &0.982$^\text{a}$&0.230$^\text{a}$&
 			0.5096$^\text{a}$&
 			0.0167$^\text{a}$\\
 			&&-34.7335$^\text{b}$& &&0.994$^\text{b}$&
 			0.263$^\text{b}$&0.5117$^\text{b}$
 			&0.1377$^\text{b}$\\
 			&&-34.7693$^{\text{c}}$&& & 0.992$^{\text{c}}$&0.227$^\text{c}$&
 			0.5000$^\text{c}$ 			& 0.0238$^\text{c}$\\
 			&&-34.7724$^{\text{d}}$& 5.7276$^\text{d}$& 0.00054$^{\text{d}}$& 0.993$^{\text{d}}$&0.342$^\text{d}$&
 			0.5000$^\text{d}$
 			& 0.0173$^\text{d}$\\[5pt]
 			\hline
 			\rule{0pt}{5ex}  7& -48.3769 &-48.1836$^\text{a}$& &&0.984$^\text{a}$&0.233$^\text{a}$&0.5077$^\text{a}$
 			&0.0207$^\text{a}$\\
 			&&-48.3181$^\text{b}$& &&0.994$^\text{b}$&
 			0.278$^\text{b}$&0.5113$^\text{b}$
 			&0.1862$^\text{b}$\\
 			&&-48.3706$^{\text{c}}$& & &0.994$^{\text{c}}$&0.234$^\text{c}$&0.4998$^\text{c}$
 			&0.0167$^\text{c}$\\
 			&&-48.3738$^{\text{d}}$&6.7512$^\text{d}$&0.00046$^\text{d}$ &0.995$^{\text{d}}$&0.352$^\text{d}$&0.5000$^\text{d}$&0.0140$^\text{d}$\\[5pt]
 			\hline
 			\rule{0pt}{5ex}8&-64.2285 &-63.9683$^{\text{a}}$& &&0.986$^\text{a}$&0.236$^\text{a}$&0.5091$^\text{a}$&
 			0.0223$^\text{a}$\\
 			&&-64.1530$^{\text{b}}$& && 0.994$^\text{b}$& 0.300$^\text{b}$&  0.5135$^{\text{b}}$
 			&0.2376$^{\text{b}}$\\
 			&&-64.2221$^{\text{c}}$&&&0.995$^{\text{c}}$&0.242$^\text{c}$& 0.5005$^\text{c}$
 			&0.0119$^\text{c}$\\
 			&&-64.2254$^{\text{d}}$& 7.7746$^\text{d}$& 0.00040$^\text{d}$&0.996$^{\text{d}}$&0.362$^\text{d}$& 0.5000$^\text{d}$
 			&0.0127$^\text{d}$\\[5pt]
 			\hline
				\rule{0pt}{5ex}   9&-82.3303 &  -81.9909$^\text{a}$&&& 0.989$^\text{a}$&0.238$^\text{a}$&0.5072$^\text{a}$ &0.0228$^\text{a}$ \\
				& &-82.2540$^{\text{b}}$&&& 0.994$^{\text{b}}$ &0.342$^\text{b}$& 0.5141$^{\text{b}}$&0.2572$^{\text{b}}$\\
				& &-82.3236$^{\text{c}}$& &&0.996$^{\text{c}}$&
				0.249$^\text{c}$&
				0.5002$^\text{c}$&0.0086$^\text{c}$\\
				& &-82.3270$^{\text{d}}$& 8.7980$^\text{d}$& 0.00038$^\text{d}$&0.997$^{\text{d}}$&
				0.373$^\text{d}$&
				0.5000$^\text{d}$&0.0122$^\text{d}$\\
				\\[5pt]
\hline
\rule{0pt}{5ex}10 &-102.6822 & -102.2504$^\text{a}$&&&0.993$^\text{a}$ & 0.240$^\text{a}$ &
				0.5064$^\text{a}$ & 0.0214$^\text{a}$\\
				& &-102.4703$^\text{b}$&&&0.996$^\text{b}$&0.374$^\text{b}$&0.5138$^\text{b}$&
				0.2186$^\text{b}$\\
				& &-102.6750$^{\text{c}}$&&&0.996$^\text{c}$&0.255$^\text{c}$&
				0.5001$^\text{c}$&0.0064$^\text{c}$\\
				& &-102.6786$^{\text{d}}$& 9.8214$^\text{d}$&0.00037$^\text{d}$
                & 0.997$^\text{d}$ & 0.384$^\text{d}$ &
				0.5000$^\text{d}$ & 0.0119$^\text{d}$\\[5pt]
\hline
\rule{0pt}{5ex}11 & -125.2842 & -124.7437$^\text{a}$ & && 0.996$^\text{a}$ & 0.241$^\text{a}$ &
				0.5051$^\text{a}$
				&0.0176$^\text{a}$\\
				& &-125.2762$^{\text{c}}$&& &0.997$^\text{c}$&0.261$^\text{c}$&
				0.5003$^\text{c}$&0.0049$^\text{c}$\\
				& &-125.2802$^{\text{d}}$ &
                10.8448$^\text{d}$&0.00037$^\text{d}$&0.998$^\text{d}$&0.395$^\text{d}$&
				0.5000$^\text{d}$&0.0118$^\text{d}$\\[5pt]
\hline
\rule{0pt}{5ex}12 & -150.1362 & -149.4590$^\text{a}$ &&& 0.999$^\text{a}$&0.242$^\text{a}$
				&0.5057$^\text{a}$
				&0.0111$^\text{a}$\\
				& &-150.1271$^{\text{c}}$& &&0.997$^\text{c}$&0.266$^\text{c}$&
				0.5005$^\text{c}$&0.0037$^\text{c}$\\
				& &-150.1317$^{\text{d}}$&11.8683$^\text{d}$&0.00038$^\text{d}$ &
                   0.998$^\text{d}$ & 0.405$^\text{d}$ &
				0.5001$^\text{d}$&0.0115$^\text{d}$\\[5pt]
\hline
\rule{0pt}{5ex}18 & -346.4987
				&-346.4616$^{\text{c}}$& & &0.998$^\text{c}$&0.289$^\text{c}$&
				0.5005$^\text{c}$
				&0.0011$^\text{c}$\\
				& &-346.4902$^{\text{d}}$& 18.0098$^\text{d}$&0.00047$^\text{d}$ &
                0.999$^\text{d}$ & 0.454$^\text{d}$ &
				0.5001$^\text{d}$ & 0.0084$^\text{d}$\\[5pt]
\hline
\rule{0pt}{5ex}20 & -429.9530&-429.8835$^{\text{c}}$&&&0.998$^\text{c}$&0.294$^\text{c}$
				&0.5001$^\text{c}$
				&0.0008$^\text{c}$\\
				& &-429.9433$^{\text{d}}$&20.0567$^\text{d}$&0.00048$^\text{d}$
                & 1.000$^\text{d}$ & 0.465$^\text{d}$ &
				0.5001$^\text{d}$ & 0.0070$^\text{d}$\\[5pt]
				\hline
 		\end{tabular}}}
\end{center}
\end{table}

\begin{table}[tb]
\caption{Ground state of Li-like atomic systems: non-linear and linear ($C, a_1, a_3, b^{(1)}_{12},
         b^{(2)}_{23}$) variational parameters {\it versus} $Z$ for compact functions {\bf (a), (c)}
         and {\bf (d)}.}
\label{params-Li}
\begin{center}
 	{\setlength{\tabcolsep}{0.6cm}
 		\begin{tabular}{ccc}
\hline
 			$\quad\quad$ Parameters & Ansatz & Fits \\
\hline
 			\rule{0pt}{4ex} $C$ &(a)& $0.036011-0.009627Z+0.003234Z^2$\\
 			&(c)&$27.3466Z^{-4}$\\
 			&(d)&$ 0.020633-0.704076Z^{-2}
 			+35.461701Z^{-4}$\\
 			\rule{0pt}{4ex} $\al_{1}^{(1)} Z$ & (a)
             &$0.118708+1.082990Z-0.007695Z^2$
 			\\
 			  &(c) &
 			  $0.132883+1.066090Z -0.001248Z^2$\\
 			  &(d) &
 			  $ 0.117679+1.075180Z-0.001314Z^2$\\
 			  \rule{0pt}{4ex} $\al_{2}^{(1)} Z$ &(a)
 			  &$-0.12485+0.762885Z+0.018467Z^2$\\
 			  &(c) &
 			  $-0.452075+0.938792Z+0.001411Z^2$\\
 			  &(d) &
 			  $-0.458461+0.931199 Z+0.001172Z^2$\\
 			  \rule{0pt}{4ex} $\al_{3}^{(1)} Z$ &(a)
 			  &$-0.601591+0.353825Z-0.004329Z^2$\\
 			  & (c) &
 			  $-0.820942+0.520317Z+0.000027Z^2$\\
 			  & (d) &
 			  $-0.794084+0.514421Z-0.000205Z^2$\\
 			  \rule{0pt}{4ex} $\al_{12}^{(1)}$ & (a)
 			  & $0.202527+0.003792Z-0.000105Z^2$\\
 			  & (c) &
 			  $0.198833+0.008703Z-0.000138Z^2$\\
              & (d) &
               $0.334012+0.000512Z-0.000006Z^2$\\
              \rule{0pt}{4ex} $c_{12}^{(1)}$   & (d)
 			  & $-0.062074 - 0.028250Z - 0.001939 Z^2$\\
 			  \rule{0pt}{4ex} $d_{12}^{(1)}$   & (d)
 			  & $-0.011397+0.004861Z+0.005260 Z^2$\\
 			  \rule{0pt}{4ex} $\al_{13}^{(1)}$ & (a) &
 			  $0.024551+ 0.004852Z+0.000059Z^2$\\
 			  & (c) &
 			  $-0.004250+0.003478Z-0.000104Z^2$\\
 			  & (d) &
 			  $0.009782-0.001745Z-0.000006Z^2$\\
 			  \rule{0pt}{4ex} $\al_{23}^{(1)}$ & (a) &
 			  $-0.033252+0.028151Z-0.000814Z^2$\\
 			  & (c) &$-0.010360+0.015041Z-0.000287Z^2$\\
 			  & (d) &$-0.010546+0.015176Z-0.000211Z^2$\\
 			\rule{0pt}{4ex} $a_{3}$ &(c)
 			&$0.254687+0.590866Z-0.007944Z^2$\\
 			&(d)&$ 0.374908 + 0.526201\,Z - 0.001442\,Z^{2}$\\
 			\rule{0pt}{4ex} $b^{(1)}_{12}$ &(d)
 			&$0.125408 - 0.033504\,Z + 0.000697\,Z^2$\\
 			\\[5pt]
 			\hline
 		\end{tabular}}
\end{center}
\end{table}

\addtocounter{table}{-1}

\begin{table}[tb]
 	\caption{ (continuation) Ground state of Li-like atomic systems: non-linear and linear
     ($a_1, b^{(2)}_{23}$) variational parameters {\it vs} $Z$ for compact functions
     {\bf (c) \rm and \bf (d)}.}
\begin{center}
 	{\setlength{\tabcolsep}{0.4cm}
 		\begin{tabular}{ccc}
\hline
 			$\quad\quad$ Parameters & Ansatz & Fits \\
\hline
 			  \rule{0pt}{4ex} $\al_{1}^{(2)} Z$ &(c)
 			  &$-0.841197+1.107970Z+0.000364Z^2$\\
 			   &(d)
 			  &$-0.662098+1.071650Z-0.004807Z^2$\\
 			  \rule{0pt}{4ex} $\al_{2}^{(2)} Z$ &(c)
 			  &$-0.152647+0.772579Z-0.000735Z^2$\\
 			   &(d)
 			  &$-0.182703+0.727755 Z + 0.008832 Z^2$\\
 			  \rule{0pt}{4ex} $\al_{3}^{(2)} Z$ &(c)
 			  &$2.027660+0.078167Z-0.002670Z^2$\\
 			  &(d)
 			  &$1.171920+0.335452Z-6\times10^{-9}Z^2$\\
 			  \rule{0pt}{4ex} $\al_{12}^{(2)}$ &(c)
 			  &$-0.154935+0.245727Z-0.002976Z^2$\\
 			  &(d)
 			  &$-0.029908+0.210006Z-0.003575Z^2$\\
 			  \rule{0pt}{4ex} $\al_{13}^{(2)}$ &(c)
 			  &$1.382690-0.430197Z+0.013276Z^2$\\
 			  &(d)
 			  &$1.129170-0.29125Z+0.001001Z^2$\\
 			  \rule{0pt}{4ex} $\al_{23}^{(2)}$ &(c)
 			  &$-0.037915+0.128535Z-0.004343Z^2$\\
              &(d)
 			  &$0.175310+0.046071Z-0.000450Z^2$\\
 			  \rule{0pt}{4ex} $c_{23}^{(2)}$ &(d)
 			  &$0.053815 - 0.001464 Z + 0.000052 Z^2$\\
 			  \rule{0pt}{4ex} $d_{23}^{(2)}$ &(d)
 			  &$-0.061936-0.001216Z+0.008384Z^2$\\
  			  \rule{0pt}{4ex} $a_1$ &(c) &
 			  $-0.484802+0.205159\,Z - 0.009918\,Z^2$\\
 			   &(d) &
 			  $-0.536084 + 0.205536\,Z - 0.002516\,Z^2$\\
 			  \rule{0pt}{4ex} $b^{(2)}_{23}$ &(d) &
 			  $-0.106216 + 0.029348\,Z - 0.000014\,Z^2$\\
 			\\[5pt]
 			\hline
 		\end{tabular}}
\end{center}
\end{table}

\begin{table}[tb]
 	\caption{Expectation values for Li-like atomic systems {\it vs}\ $Z$ using the function
             {\bf (d)}; \\ ${}^{(a)}$ \cite{Yan-Drake:1995}, ${}^{(b)}$ \cite{TGH2009},
             ${}^{(c)}$ the function {\bf (c)}, see text
             }
\label{exp-val-Li}
\begin{center}
 	{\setlength{\tabcolsep}{0.4cm}
 		\begin{tabular}{ccc}
\hline
 			$Z$ & $\langle r_{ij} \rangle$ & $\langle r^{-1}_{ij}\rangle $\\
\hline
3    &   8.6711          & 2.2037        \\
     &\   8.6684 $^{a}$   &\ 2.1982 $^{a}$ \\
     &\   8.6552 $^{b}$   &\ 2.2091 $^{b}$ \\
     &\   8.6531 $^{c}$   &\ 2.2143 $^{c}$ \\
4    &	 5.2771   &     3.2511  \\
5    & 	 3.8383	  &     4.2853  \\
10   &	 1.6554   &     9.3997	\\
     &\	  1.6520 $^{c}$   &\ 9.4161 $^{c}$ \\
12   &	 1.3466   &    11.4551  \\
18   &	 0.8681   &    17.5826	\\
20   &	 0.7764   &    19.6227	\\
     &\	  0.7717 $^{c}$   &\ 19.6467 $^{c}$ \\
\hline \hline
 		\end{tabular}}
\end{center}
\end{table}

\vspace{-8mm}
\section{Conclusions}
\vspace{-5mm}

For the ground state of the He-like two-electronic sequence $(Z,2e)$, which is spin-singlet for any $Z$, two ultra-compact accurate symmetric functions $\Psi^{(+)}$ are constructed: (i) a straightforward 5-parametric generalization of the Hylleraas and Kinoshita functions $\Psi^{(+)}_{HK}$, which allow calculation of variational energy integrals and polynomial expectation values in closed analytic form, and (ii) a non-trivial 7-parametric generalization $\Psi^{(+)}_F$ of function (i), which allows one to get the same relative accuracy in both cusp parameters and electronic correlation energy. Both functions allow to get absolute accuracy in energy of 3 d.d. for $Z \in [1,20]$, thus, they describe the domain of applicability of QMCC.
Needless to say both functions $\Psi^{(+)}_{HK,F}$ can be used as entries for the non-linearization procedure \cite{Turbiner:1980-4} to increase accuracy in energy and wavefunction with sufficiently high rate of convergence.

The function $\Psi^{(+)}_{HK}$ (i) provides much higher accuracy than the Hylleraas function. It can be used to form linear combinations as trial functions leading presumably to faster convergence than the ones which were based on Hylleraas functions \cite{Korobov:2000} preserving the property to calculate all involved integrals analytically.
Function (ii) appears as uniform, locally accurate approximation of the exact ground state eigenfunction: it provides the same (or better) relative accuracies in energies and several expectation values (which are among the most difficult to get accurate results for) as well as both cusp parameters for all studied values of $Z \in [1,20]$. The parameters of function (ii) are smooth, easily fitted functions of $Z$. It has to be mentioned that both functions $\Psi^{(+)}_{HK,F}$ are characterized by the existence of a nodal surface, where $\Psi^{(+)}_{HK,F}=0$, e.g. if $r_1=r_2=r$, it is $(1 - a r + b r_{12})=0$. Surprisingly, it still allows us to get highly accurate results in energy and expectation values for $Z\in[1,20]$. This nodal surface continues to exist even if all constraints are relaxed and straightforward variational calculations are carried out.
Evidently, for the spin-triplet state with the lowest energy, the ultra-compact function can be of a form similar to (\ref{psinew}) but antisymmetric,
\begin{equation*}
   \Psi^{(-)}_F\ =\ (1-P_{12})\,(1 - a r_1 + b r_{12})\,e^{- \al Z r_1\, -\, \beta Z r_2\, +\, \gamma r_{12}\, \frac{(1 + c r_{12})}{(1 + d r_{12})}}\ ,
\end{equation*}
of course, with different parameters $(a, b, \al, \beta, \gamma, c, d)$.
The results for this state as well as for a few spin-singlet and spin-triplet excited states of Helium-like sequence will be presented in forthcoming work \cite{Part-2:2020}.

For the ground state of Li-like sequence $(Z,3e)$, which is spin-doublet for all $Z$,
four ultra-compact functions {\bf (a), (b), (c), (d)} are constructed in the form
of a generalization of the previous ones. They depend on 7, 13, 15 and, eventually, 21 variational parameters, respectively, all of which admit physical interpretation. With the increase in the number of parameters, the accuracy for the energy and the electron-nuclear cusp grows systematically for all $Z\in[3,20]$: as for the energy of the 21-parametric ultra-compact {\bf (d)}: (\ref{Phi-1-mod}),(\ref{Phi-2-mod}),(\ref{Phi-d}), the absolute deviation changes monotonously from $\sim 0.0024$\,a.u. at $Z=3$ to $\sim 0.01$\,a.u., while for the (normalized) electron-nuclear cusp the relative deviation changes monotonously from $\sim 1.2\%$ to $\sim 0.1\%$. Hence, the domain of applicability of QMCC for energy is described. Contrary to that, the electron-electron cusp parameter does not display the systematic improvement with the number of parameters for the ultra-compact functions used. The most accurate results are obtained with the 21-parametric ultra-compact {\bf (d)}: the electron-electron cusp $C_{e,e}$ grows monotonously from $0.326$ at $Z=3$ to $0.465$ at $Z=20$. Thus, it deviates from $\sim 37\%$ to $\sim 7\%$ from the exact electron-electron cusp depending on $Z$. The possibility of constructing ultra-compact, uniform, locally accurate function based on the 21-parametric function {\bf (d)}: (\ref{Phi-1-mod}),(\ref{Phi-2-mod}),(\ref{Phi-d}), which supports the relation between the electron-nuclear and the electron-electron cusps
\[
  {C}_{Z,e}\ =\ 2\, Z\, C_{e,e}\ ,
\]
similar to the one for the He-like sequence, see (\ref{1 to 2}), will be discussed in the forthcoming paper \cite{Part-2:2020}.
For the 21-parametric function {\bf (d)} in domain $Z \in [3,20]$ all 20 parameters are fitted by second degree polynomials in $Z$ with a small coefficients in front of $Z^2$ when the remaining linear parameter $C$ is a second degree polynomials in $1/Z^2$. The ground state energy is well approximated by the Majorana polynomial in $Z$, see (\ref{majorana-3}). It is demonstrated that the probability of finding one of two spin 1/2 functions $\chi_{1,2}$ is equal to $1/2$.

Spin-doublet function {\bf (d)}, see (\ref{Phi-1-mod}),(\ref{Phi-2-mod}),(\ref{Phi-d}), is a linear superposition of two spin 1/2 functions. It can be easily modified to the one for a spin-quartet state with single spin function $\al\al\al$. The results for spin-quartet state of the lowest energy will be presented in the forthcoming paper \cite{Part-2:2020}.

\vspace{-5mm}
\section*{Acknowledgments}
\vspace{-3mm}

The authors thank H.~Olivares~Pil\'on for their interest to the work and useful discussions. J.C.~del\,V. is supported by CONACyT PhD Grant No.570617 (Mexico). This work is partially supported by CONACyT grant A1-S-17364 and DGAPA grant IN113819 (Mexico).

\appendix
\section{Two-electron system with fixed charge $Z$}
\label{appendix}

Take the function (\ref{psinew})
\begin{equation*}
   \Psi_F\ =\
   \bigg[(1 - a Z r_1 + b r_{12})\,e^{- \al Z r_1\, -\, \beta Z r_2\, +\, \gamma r_{12}\,
   \frac{(1 + c r_{12})}{(1 + d r_{12})}}\ +\ (1 \lrar 2)\bigg]\ ,
\end{equation*}
where $a, b, c, d, \al, \beta, \gamma$ are parameters.

\bigskip

\noindent
{\bf OBSERVATION-I\ (E.~Hylleraas, 1929).} \quad At $Z \rar \infty$ the Helium atom is reduced to two $Z$-Hydrogen atoms, the Hamiltonian (\ref{H-He}) becomes the sum of two Hydrogenic Hamiltonians,
\[
    {\cal H}\ =\ {\cal H}_1 + {\cal H}_2\ \equiv \
    -\left(\frac{1}{2}\, \De_1 \ +\ \frac{Z}{r_1}\right)\ -\ \left(\frac{1}{2}\, \De_2\ +\ \frac{Z}{r_2}\right)\ ,
\]
its ground state function (\ref{He-infty-psi}) coincides with (\ref{psinew}) at $a=b=c=d=\gamma=0$ and $\al=\beta=1$,
\begin{equation*}
   \Psi_F\ =\
   \bigg( e^{- Z r_1\, -\, Z r_2}\ +\ (1 \lrar 2)\bigg)\ ,
\end{equation*}
hence, it is known exactly, $E_F=-Z^2$ (\ref{He-infty-E}).

\bigskip

\noindent
{\bf Proof.} \ By direct calculation.

\bigskip

\noindent
{\bf OBSERVATION-II.}\quad
For $Z \in [1,20]$, if the parameters in (\ref{psinew}) are chosen to be
\begin{align}
\label{parameters}
 \al_F    &\ =\ -\,\frac{0.0673}{Z}\ +\  1.049647\ -\ 0.000842\, Z\ , \non \\
 \beta_F  &\ =\ -\, \frac{0.466383}{Z}\ +\ 0.919016\ +\ 0.001578\,Z\ , \non \\
 \gamma_F &\ =\ -\,\frac{0.704459\ +\ 0.752684\,Z\ -\ 1.52103\,Z^2}
                 {4.278577\ -\ 4.017493\,Z\ -\ 0.388753\,Z^2}\ , \non \\
 a_F      &\ =\ (4 c_{e,e}\ -\ \al_F\ -\ \beta_F) \ , \\
 b_F      &\ =\ c_{e,e}\ -\ \gamma_F \ , \non \\
 c_F      &\ =\  -\,\frac{0.036094}{Z}\ +\ 0.035756\ -\ 0.007766\,Z\ -\ 0.001361\,Z^2 \ ,  \non \\
 d_F      &\ =\  -\,\frac{0.090021}{Z}\ +\ 0.286716\ +\ 0.040912\,Z\ -\ 0.002569\, Z^2 \ , \non
\end{align}
with
\begin{equation}
\label{cee}
  c_{e,e} \ =\ \frac{1}{2}\ -\ \frac{0.05}{Z}\ +\ \frac{0.024}{Z^2}\ -\ \frac{0.021}{Z^3} \ ,
\end{equation}
cf. Fig.1 for illustration, then in $(r_1,r_2,r_{12})$ space the relative deviation from the exact ground state function $\Psi_0$ is bounded so that
\[
   \bigg| \frac{\Psi_0 - \Psi_F}{\Psi_0} \bigg| \ \leq \ (1 - 2 c_{e,e})\ ,\qquad (\star)
\]
as well as the relative deviation from exact ground state energy $E_0$ wrt correlation energy
\[
      \bigg| \frac{E_0 - E_M}{E_0 + Z^2} \bigg| \ \leq\ 0.02 \times (1 - 2\,c_{e,e})\ ,\qquad (\star\star)
\]
where
\begin{equation*}
\label{Majorana}
     E_M\ =\ -\,0.153282\ +\ 0.624583\, Z\ -\,Z^2\ ,
\end{equation*}
cf.(\ref{majorana-2}), is the approximation of the ground state energy via Majorana polynomial \footnote{This formula was called in \cite{AOP:2019} the Majorana formula, it holds for all $Z \geq 1$ approximating the lowest eigenvalue with accuracy 4 figures for $Z \leq 10$ and 5 figures (or more) for $Z \geq 11$. }
. For any expectation value $\langle\, O\, \rangle|_{\Psi_0}$
the relative deviation,
\[
   \bigg| 1\ -\ \frac{\langle\, O\, \rangle |_{\Psi_F}}{\langle\, O\, \rangle|_{\Psi_0}} \bigg| \ \leq \ (1 - 2 c_{e,e})\ ,\qquad (\star\star\star)
\]
as well as for the cusp parameters $C$, which appear to be equal,
$\frac{C_{Z,e}}{Z}=2C_{e,e}$,
\[
   \bigg| 1\ -\frac{C_{Z,e}}{Z} \bigg| \ \leq \ (1 - 2\,c_{e,e})\ ,\
   \bigg| 1\ - C_{e,e} \bigg| \ \leq \ (1 - 2\,c_{e,e})\
   .\qquad (\star\star \star \star)
\]

\noindent
{\bf Demonstration.} By making comparison with available accurate numerical results one can see that Observation $(\star\star)$ holds for $Z\in[1,20]$; $(\star\star\star)$ holds for the nine expectation values, see below, for $Z \in [1,10]$, which are the most difficult (from the technical point of view) to calculate with high precision,
\[
   \langle H_0 \rangle \ ,\
   \langle\bf{r}_1\cdot\bf{r}_2\rangle \ ,
\]
\[
  \langle \de({\bf r}_{1,2}) \rangle\ ,\ \langle \de({\bf r}_{12}) \rangle\ ,\ \langle\delta(\bf{r}_1)\de(\bf{r}_2)\rangle
\]
\[
   \langle \de({\bf r}_{1,2})\frac{\pa}{\pa r_{1,2}} \rangle\ ,\ \langle \de({\bf r}_{12})\frac{\pa}{\pa r_{12}} \rangle\ ,
\]
and $(\star\star\star\star)$ holds for all studied $Z$. The results are stable wrt small variations of parameters (\ref{parameters}) and (\ref{cee}).

To demonstrate the validity of $(\star)$ it has to be analysed using 3-dimensional electrostatics with variable permitivity given by $\Psi_F^2$. It is described by elliptic PDE in 3 variables with the first derivatives present in $(r_1,r_2,r_{12})$ space. This will be done elsewhere.

%
%
%
%
%
%
%
%
 \newpage

\section*{Plots of parameters (\ref{parameters}), (\ref{cee}) for $Z \in [1,20]$}

  \begin{figure}[hb]
\centering
\subfloat[]{
  \includegraphics[width=75mm]{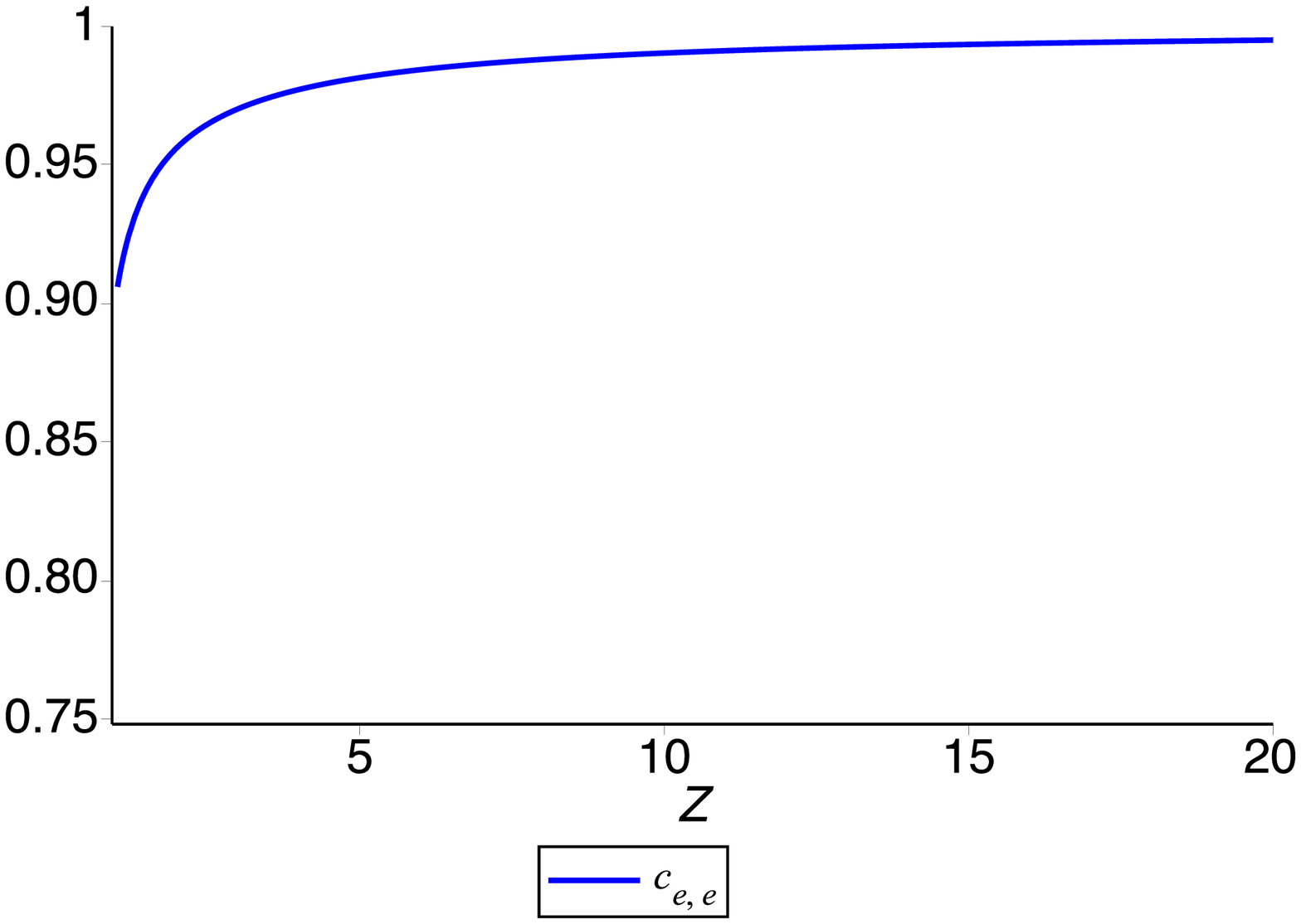}
}
\subfloat[]{
  \includegraphics[width=75mm]{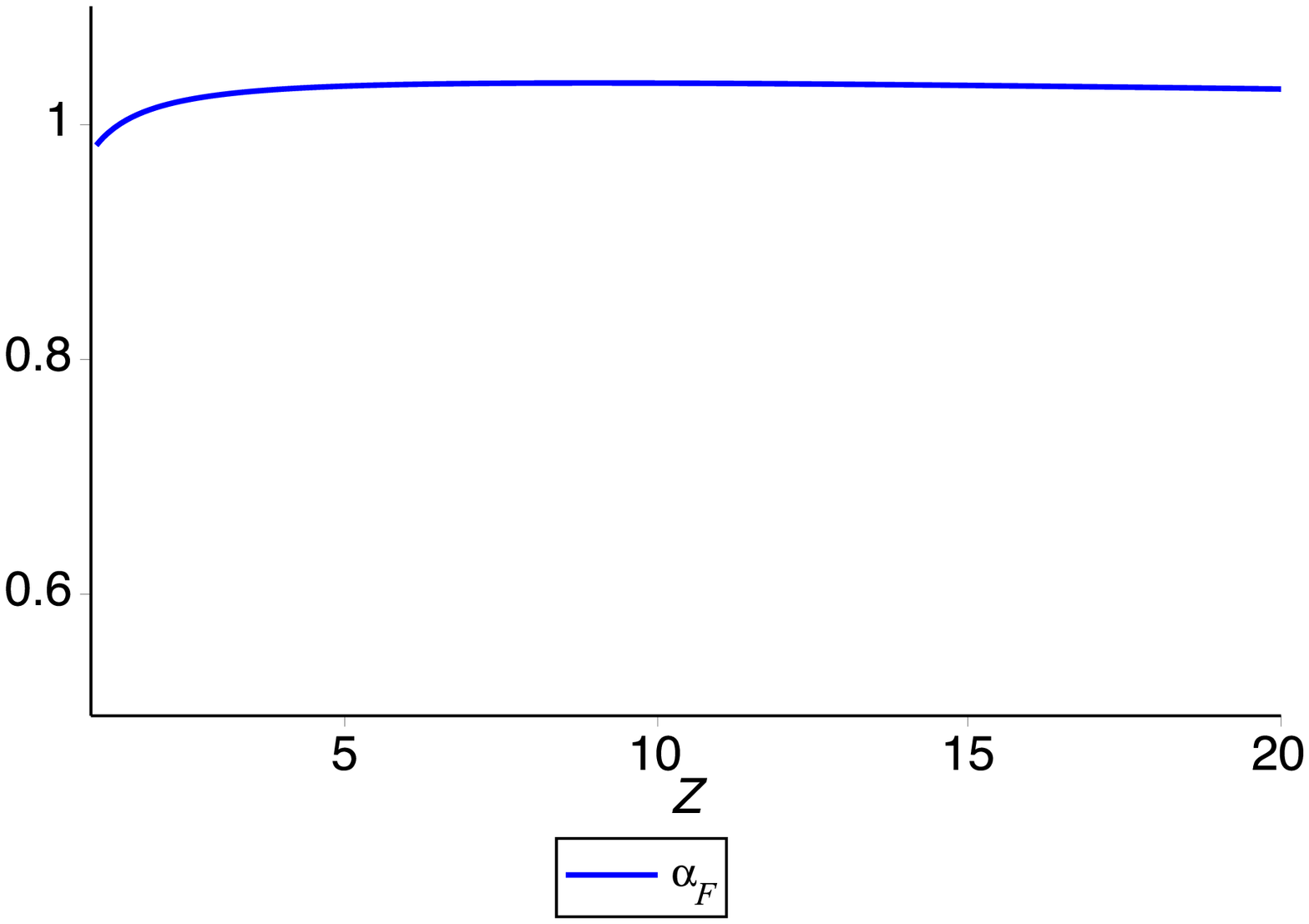}
}
\hspace{0mm}
\subfloat[]{
  \includegraphics[width=75mm]{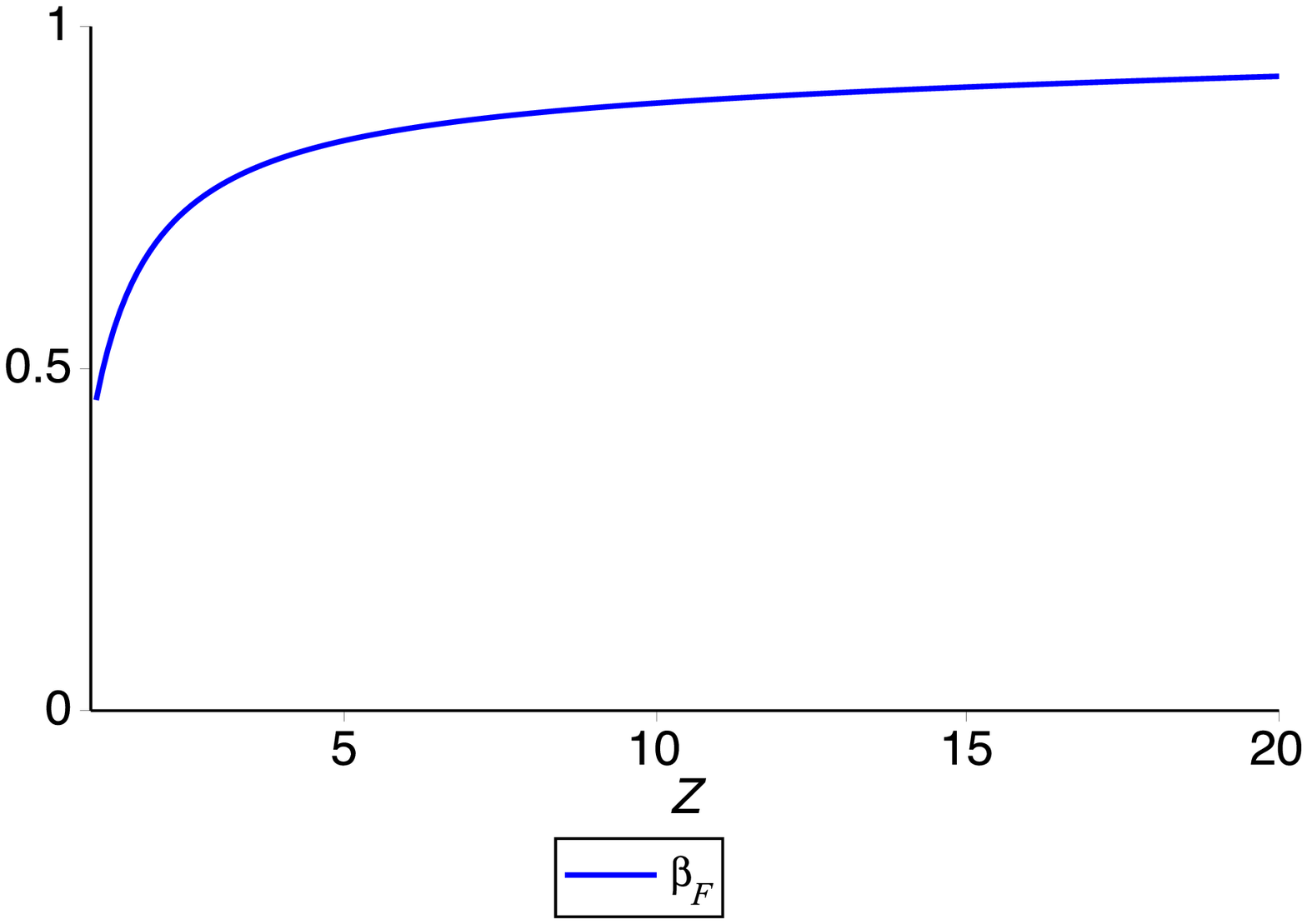}
}
\subfloat[]{
  \includegraphics[width=75mm]{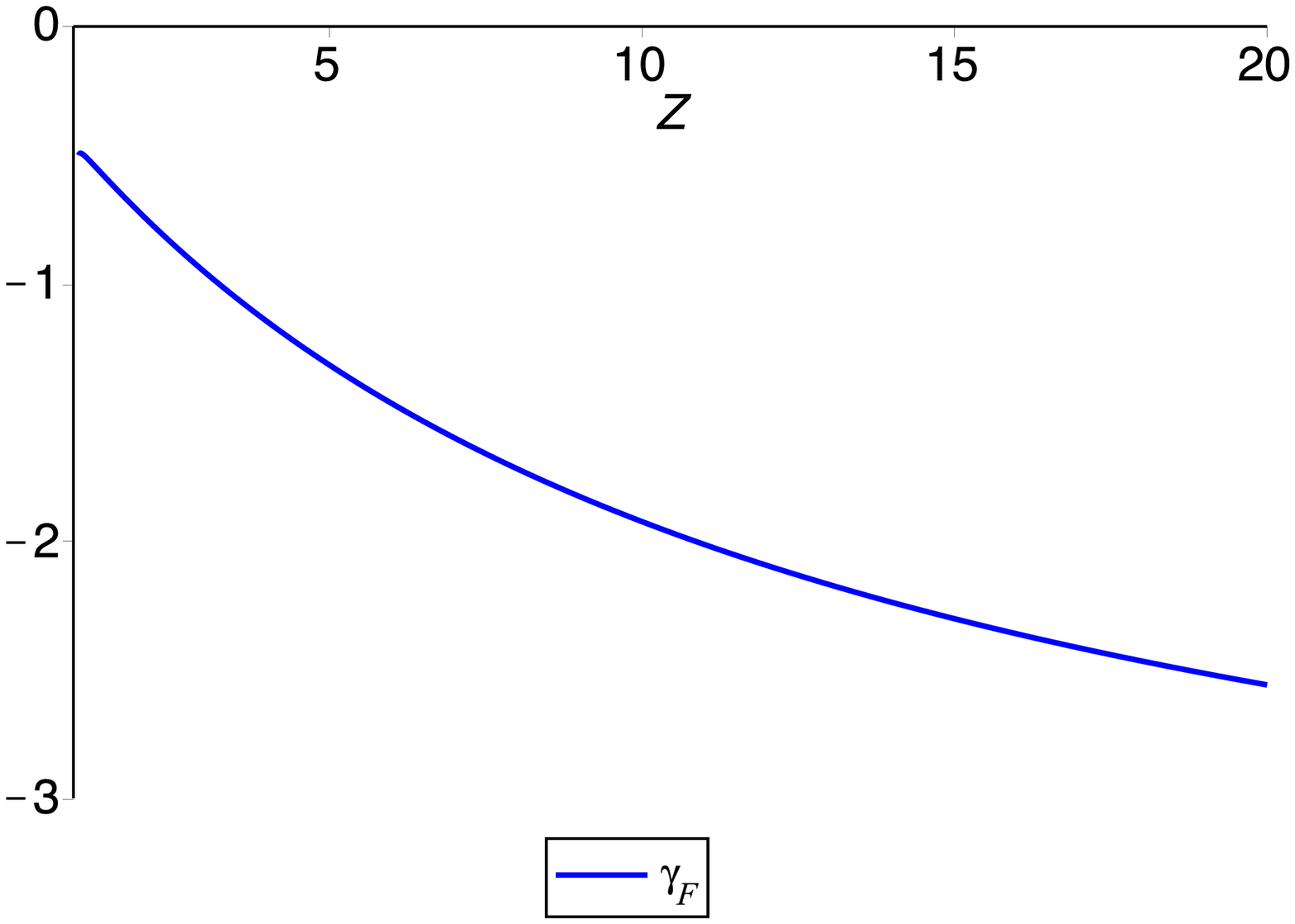}
}
\caption{Parameters $\al_F, \beta_F, \gamma_F$ (\ref{parameters}) and $c_{e,e}$ (\ref{cee}) {\it vs}\ $Z$; cusp parameters $\frac{C^{(F)}_{Z,e}}{Z}=2C^{(F)}_{e,e} = c_{e,e}$}
\end{figure}
 \begin{figure}
 \ContinuedFloat
\centering
\subfloat[]{
  \includegraphics[width=75mm]{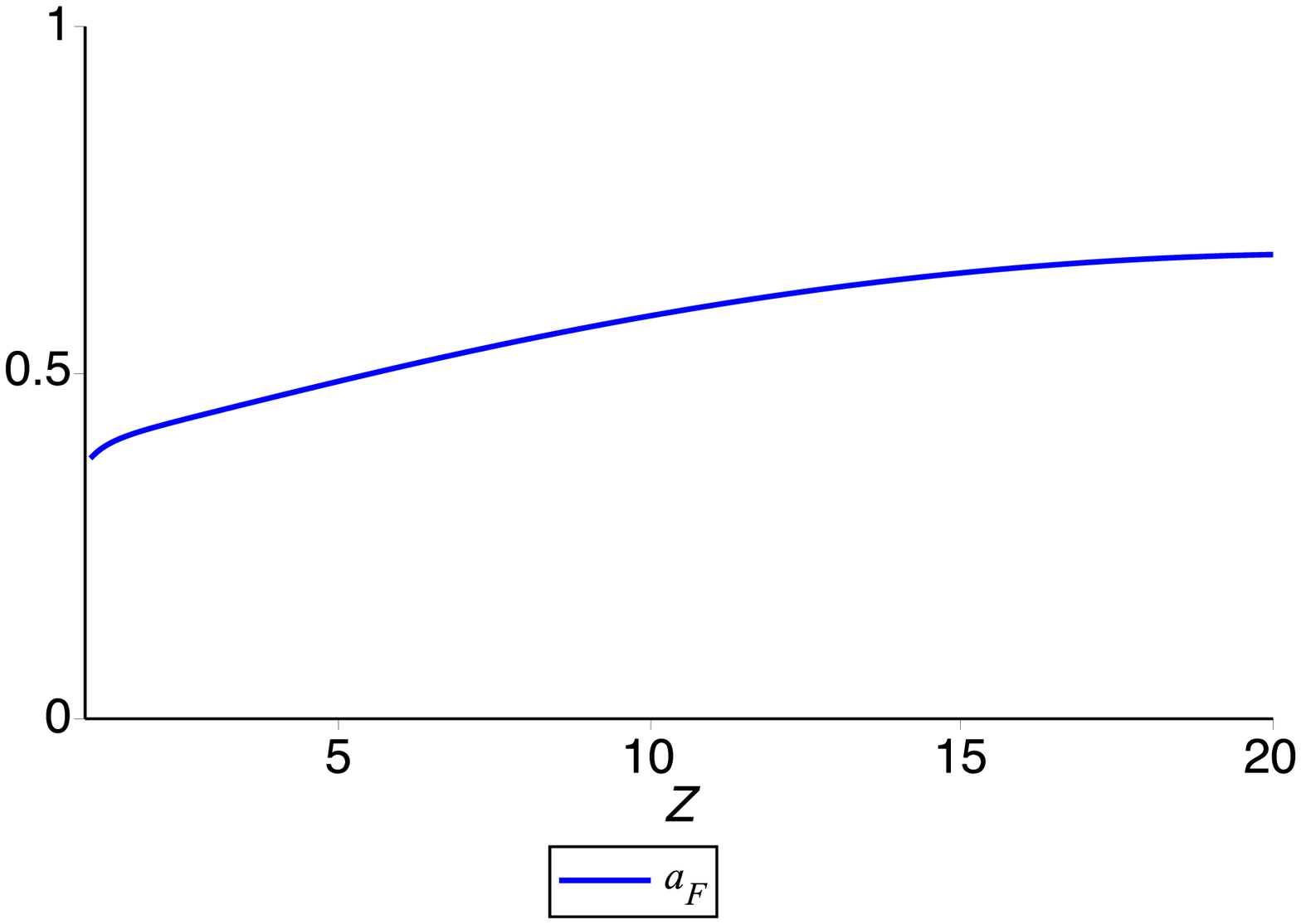}
}
\subfloat[]{
  \includegraphics[width=75mm]{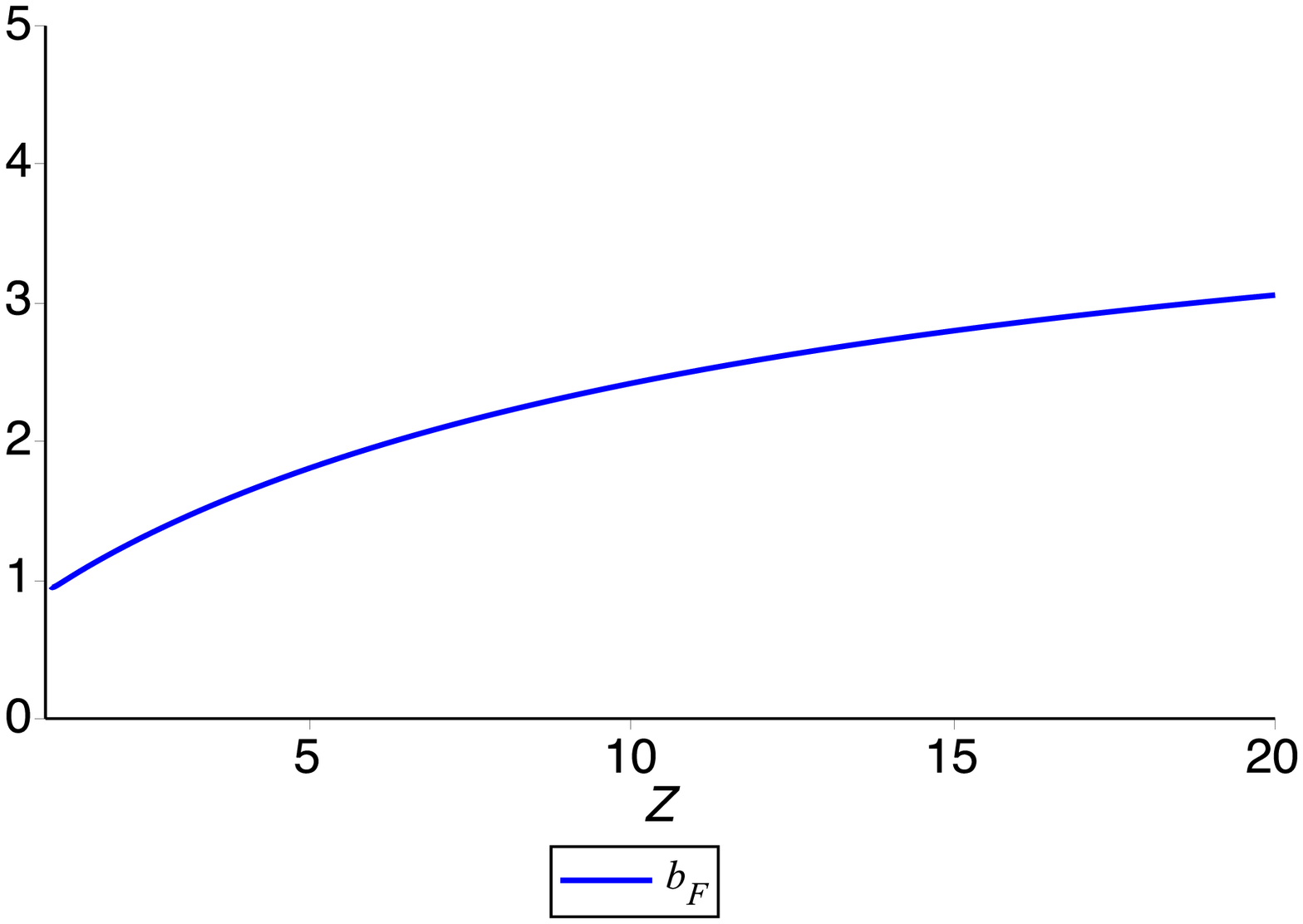}
}
\hspace{0mm}
\subfloat[]{
  \includegraphics[width=75mm]{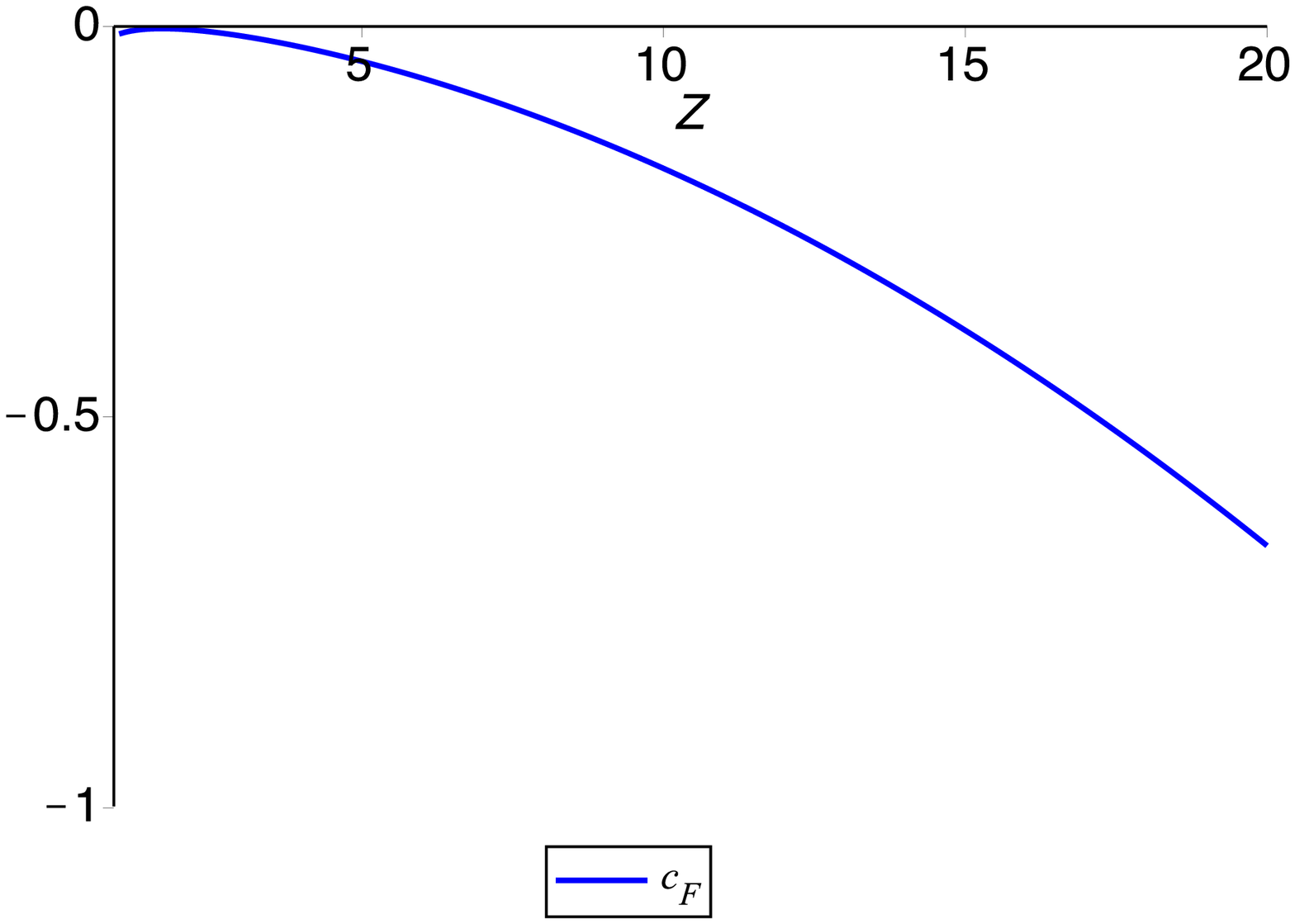}
}
\subfloat[]{
  \includegraphics[width=75mm]{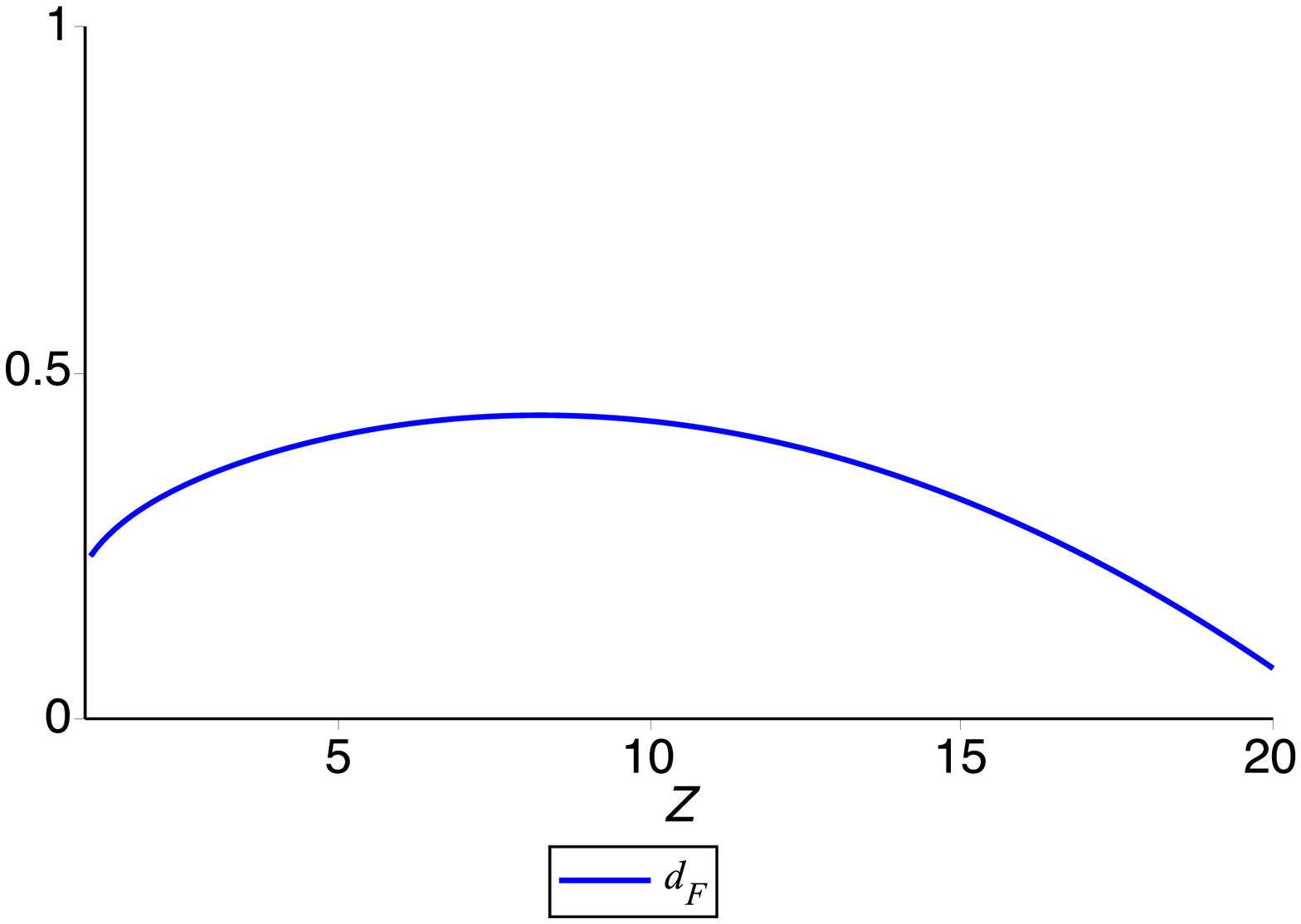}
}
\caption{ (continuation)\ Parameters $a_F, b_F, c_F, d_F$ (\ref{parameters}) {\it vs}\ $Z$}
\end{figure}

\newpage



\begin{thebibliography}{99}

\bibitem{Turbiner:1980-4}
          A.~V.~Turbiner,\\
           {\it Soviet Phys. -- ZhETF \bf 79}, 1719-1745 (1980),\\
           {\it JETP \bf 52}, 868-876 (1980)
           (English Translation);\\
          {\it Soviet Phys. - Usp. Fiz. Nauk. \bf 144}, 35-78 (1984),\\
          {\it Sov. Phys. Uspekhi \bf 27}, 668-694 (1984) (English Translation)

\bibitem{Hylleraas}
        E.A.~Hylleraas, \\
        {\it Neue Berechnung der Energie des Heliums im Grundzustande,
        sowie des tiefsten Terms von Ortho-Helium},\\
        {\em Z.~Phys. \bf 54} 347-366 (1929) (in German);\newpage
        English translation: Quantum chemistry: classic scientific papers, translated and edited
        by H Hettema,  Singapore; London:  World Scientific, 2000, pp 104-121.

\bibitem{AOP:2019}
        A.V.~Turbiner, J.C.~ L\'opez Vieyra,  H.~Olivares Pil\'on,\\
        {\it Annals of Physics \bf 409} (2019) 167908 (19 pp)

\bibitem{Berezin-Shubin}
          Felix~A.~Berezin and Mikhail~A.~Shubin,\\
          {\it The Schr\"odinger Equation}\\
          Publisher: Springer Science \& Business Media, 1991, pp. 555 \\
          ISBN	079231218X, 9780792312185	

\bibitem{Harris}
         F.E.~Harris and V.H.~Smith Jr.,\\
         {\it J. Phys. Chem. \bf A 109} (2005) 11413-11416

\bibitem{Mulliken}
        R.S.~Mulliken, \\
        {\it  J.~Chem.~Phys. \bf{43} \rm{S2}} (1965)

\bibitem{Carter:1966}
          B.P.~Carter,\\
          {\em Phys. Rev. \bf 141} (1966) 863-872

\bibitem{Korobov:2000}
          V.I.~Korobov,\\
         {\em Phys. Rev. \bf A 61} (2000) 064503;\\
         {\em Phys. Rev. \bf A 66} (2002) 024501;\\
          D.T.~Aznabaev, A.K.~Bekbaev, and V.I.~Korobov,\\
          {\em Phys. Rev. \bf A 98} (2018) 012510

\bibitem{delValle}
          J.C.~Valle, A.~V.~Turbiner,\\
          {\it Int.Journ.Mod.Phys. \bf A34} (2019) 1950143 (43pp);\\
          {\it Int.Journ.Mod.Phys. \bf A35} (2020) 1950143 (45pp)

\bibitem{TGH2009}
          N.L.~Guevara, F.E.~Harris and A.V.~Turbiner,\\
          {\it Int. Journ. Quant. Chem,  \bf 109, \rm 3036-3040 (2009)}

\bibitem{twe}
   A.V.~Turbiner, W.~Miller Jr and M.A.~Escobar Ruiz,\\
   {\it Journal of Physics \bf A50} (2017) 215201;\\
   {\it Journ of Math Physics \bf A59} (2018) 022108;\\
   {\it Journal of Physics \bf A51} (2018) 205201;\\
   {\it Journ of Math Physics \bf A60} (2019) 062101

\bibitem{GAM:1987}
         J.E.~Gottschalk et al,\\
         {\it J Phys \bf A20} (1987) 2077-104

\bibitem{Myers:1991}
        C.R.~Myers et al.\\
        {\it Phys Rev \bf A44} (1991) 5537-46

\bibitem{Bartlett:1937}
         J.H.~ Bartlett,\\
         {\it Phys Rev \bf 51} (1937) 661-9

\bibitem{CS:1969}
         D.P.~Chong and D.M.~Schrader,\\
         {\it Molecular Physics \bf 16} (1969) 137-144


\bibitem{CL}
        J.-L.~Calais and P.-O.~L\"owdin,\\
        {\em J. Mol. Spectr. \bf 8}, 203-211 (1962)

\bibitem{Kinoshita}
        T.~Kinoshita, \\
        {\it Phys. Rev. \bf 105}, 1490-1502  (1957)

\bibitem{Drake}
        G.~W.~F.~Drake,\\
        in \textit{Springer Handbook of Atomic, Molecular, and Optical Physics}\\
        (Ed: G.W.F.~Drake), Springer New York, New York,  Ch. 11, pp. 199-219 (2006).

\bibitem{Frolov}
        A.M.~Frolov and V.H.~Smith~Jr.,\\
        {\em J. Phys. \bf B37}, 2917-2932 (2004)

\bibitem{ThakkarSmith:77}
         A. J. Thakkar and V.H.~Smith~Jr.,\\
         {\it Phys. Rev. \bf A 15}, 1-15  (1977)

\bibitem{Pekeris:58}
              C. L. Pekeris,\\
               {\it Phys. Rev. \bf 112}, 1649-1658  (1958)

\bibitem{Nakashima:2008}
              H. Nakashima and H. Nakatsuji, \\
              {\it J. Chem. Phys. \bf 128} 154107 (2008)

\bibitem{Pauncz}
        R. Pauncz,\\
        {\it Spin eigenfunctions – Construction and use}
        {\rm  (Plenum Press, New York, 1979).}

\bibitem{Frolov:2010}
        A.M.~Frolov and D.M. Wardlaw,\\
        {\it JETP} 138, 5 (2010)

\bibitem{Larsson}
        S.~Larsson,\\
        {\it Phys. Rev. \bf 169}, 49 (1968)

\bibitem{BMBM:2001}
        L.~Bertini, M.~Mella, D.~Bressanini, G.~Morosi,\\
        {\em J. Phys. \bf B34}, 257-265 (2001)

\bibitem{Puchalski}
   M. Puchalski and K. Pachucki,\\
   {\it Phys. Rev. A} {bf 73}, 022503 (2006)

\bibitem{Fromm-Hill}
          D.M.~Fromm, R.N.~Hill,\\
          {\em Phys. Rev. \bf A 36}, 1013-1044 (1987)

\bibitem{Yan-Drake:1995}
          Z.-C.~Yan and G.W.F.~Drake,\\
          {\it  Phys. Rev. \bf A 52}, 3711 (1995)

\bibitem{Part-2:2020}
        D.J.~Nader, J.C.~Valle, J.C.~ L\'opez Vieyra, A.V.~Turbiner,\\
        {\it Ultra-Compact accurate wave functions for He-like and Li-like iso-electronic sequences and variational calculus. II. Lowest excited states}\\
        (in progress)

\end{thebibliography}
\end{document}